\RequirePackage{amsmath}
\documentclass{aa}
\usepackage{mathtools}

\usepackage{graphicx,natbib}
\usepackage[latin1]{inputenc}
\usepackage[T1]{fontenc}
\usepackage{epstopdf}
\bibpunct{(}{)}{;}{a}{}{,}
\def\arxivprefixesep{:}

\newcommand{\eprint}[2][]{%
{\texttt{\if!#1!#2\else#1\arxivprefixesep\ignorespaces#2\fi}}%
}
\usepackage{txfonts}
\usepackage{bm}
\usepackage{braket}
\usepackage{tikz}
\usetikzlibrary{positioning}

\usepackage[separate-uncertainty]{siunitx}
\DeclareSIUnit{\mearth}{M_\oplus}
\DeclareSIUnit \au {AU}
\DeclareSIUnit \umass {M_{\odot}}
\DeclareSIUnit \yr {yr}
\usepackage{subfigure}
\usepackage{color}

\let\bibfont\relax
\begin{document}

\title{Radial Drift and Concurrent Ablation of Boulder-Sized Objects}

\author{Remo Burn, Ulysse Marboeuf, Yann Alibert \& Willy Benz}
\institute{Physikalisches Institut  \& Center for Space and Habitability, Universit\"at Bern, CH-3012 Bern, Switzerland,
        \email{remo.burn@space.unibe.ch}
}

\abstract
{The composition of a protoplanetary disk at a given location does not only depend on temperature and pressure but also on the time dependent transport of matter, such as radial drift of solid bodies, which could release water and other volatile species before disintegration or accretion onto a larger body with potentially considerable implications for the composition of planets.}
{We perform a parameter study focused on the water depletion of different sized bodies able to cross the water snowline by gas induced radial drift.}
{Either the analytical Hertz-Knudsen-Langmuir sublimation formula assuming equilibrium temperature within the body or a more involved, numerical model for the internal thermal evolution is coupled with an $\alpha$-disk model. Different properties of the disk and the embedded body are explored.}
{Bodies with radii up to \SI{100}{\meter} drift faster towards the central star than the water snowline, hence, cross it. The region that can be reached before complete disintegration -- and is therefore polluted with H$_2$O ice -- extends to \SI{10}{\percent} closer to the star than the snowline location. The extent of this polluted region could be multiple times larger in the presence of a dust mantle, which is, however, unlikely to form due to frequent collisions with smaller-than centimeter sized objects.}
{Given a significant abundance of meter sized boulders in protoplanetary disks, the transport of water by radial drift of these bodies towards regions closer to the star than the snowline is not negligible and this flux of volatiles can be estimated for a given distribution of solid body sizes and compositions. A simple expression for surface sublimation is applicable for a homogeneous body consisting of only dust and water ice without a dust mantle.}

\authorrunning{R. Burn et al.}
\titlerunning{Radial Drift and Concurrent Ablation of Boulder-Sized Objects}
\maketitle

\section{Introduction}
Recent years of observations and theoretical work on planet formation have stressed the importance of the physics at the various snow- or icelines, i.e. the regions in the protoplanetary disk where a volatile species reaches its condensation temperature. Rather than defining snowlines using the condensation temperature, it is more relevant for planet formation to focus on the presence of water ice in building blocks of planets, which might be different due to dynamical processes. However, we will keep the notion of snowline to refer to the "classic" snowlines based on temperatures and pressures only.\footnote{Here, we generally ignore the fact that a snowline is a surface with a strong dependence on height, but instead only consider the snowline position at the midplane.}

Thanks to the continuous improvement of radio-astronomical facilities, such as the Atacama Large Millimeter/Submillimeter Array (ALMA), observations of the carbon monoxide (CO) snowline in certain disks are now possible \citep{Qi2013,Qi2015, Schwarz2016, Nomura2016, Guidi2016}. The CO snowline is the most accessible snowline to observation because of the low freezing point (\SI{30}{\kelvin}-\SI{40}{\kelvin} \citep{Fray2009}), implying a large distance to the star, and the high abundance of CO in the disk gas. Unfortunately, the H$_2$O snowline is harder to observe owing to the higher condensation temperature of water. So far, observations were limited to a disk heated by a stellar outburst \citep{Cieza2016}.

The main interest on the water snowline stems from emerging compositional studies of terrestrial planets, motivated by increased precision on measured radii and masses of planets using radial velocity measurements combined with \textit{Kepler} transit data \citep[e.g.][]{Marcy2014}, or transit timing variation \citep[e.g.][for the TRAPPIST-1 system]{Gillon2017, Grimm2018}. Theoretical models of planet formation may help break the degeneracy between planets covered by oceans and those containing H, He atmospheres, while also constraining the mantle composition, if they can put reliable constraints on the volatile content of planets \citep{Adams2008, Rogers2010, Dorn2015}. To achieve this, the compositions of solids and gas in the disk, which are accreted by the (migrating) planets, have to be well constrained over a large region. This will be ultimately necessary to assess the habitability of observed exoplanets.

Finally, the recent studies of \citet{Ida2016,Drazkowska2017,Schoonenberg2017} show increased planetesimal formation rates by streaming instability at the H$_2$O snowline, the latter two taking released water vapor into account \citep[see also][]{Ros2013}. Those results show the need for proper treatment of all occurring physical processes at the snowline.

A redefinition of the snowline for asteroids in the solar system was explored by \cite{Schorghofer2008} using similar means to what we will present. \cite{Schorghofer2008} found that ice can persist on asteroids of kilometer size up to temperatures of at least \SI{145}{\kelvin} over the solar system lifetime and calculated for multiple parameters where these conditions occur. Other recent works aimed at determining the disk composition used chemically evolved \citep{Visser2009,Eistrup2016} or equilibrium chemistry \citep{Thiabaud2015} disks as a basis. However, in these works, no transport of solids, such as radial drift \citep{Weidenschilling1977} or diffusion processes were modeled.
In addition to the relevant chemical evolution, these dynamical effects need to be considered \citep[see][for a recent review]{Pudritz2018}. Some studies that do address radial transport of different species by pebbles and vapor are \citet{Stevenson1988, Drazkowska2017, Schoonenberg2017}. Here, we will investigate the potential impact of boulder-sized bodies (sizes from \SI{}{\centi\meter} to \SI{100}{\meter}), which is a size regime not treated in the aforementioned works. Such bodies might efficiently transport water ice to regions starwards of the classical water snowline by drifting through the disk faster than the snowline is moving towards the star. In general, this fast drift leads to fast removal of these bodies, which is usually used as an argument not to include those sizes in models. Furthermore, coagulation processes are not efficient enough to let a body grow directly to this size range under nominal conditions \citep[see][for a recent review]{Blum2018}. However, bodies in this range of sizes are present in the current asteroid population \citep[e.g.][]{Bottke2005} and models suggest they are naturally produced in collisions of larger bodies \citep[e.g][]{Benz1999, Bottke2015}.

In this work, we postulate the presence of meter-sized objects and perform a parameter study to determine the region that can be reached by fast drifting bodies crossing the water snowline. This will let us gauge the importance of modeling solid ice transport of fast drifting bodies in the disk. For this, we will identify the dominating processes contributing to thermal heating of the bodies and the parameters and properties influencing the process. Furthermore, we test a simplified, analytic model for the sublimation of water on a boulder-sized body against a more complex, cometary nucleus model \citep{Marboeuf2012} adjusted to account for the presence of protoplanetary disk gas in the vicinity. The application of a simple analytical expression instead of a more complex numerical model for the sublimation of water ice would help to substantially reduce the computational cost and complexity of future works.

In Sect. \ref{sec:model}, we describe the different parts and modes (numerical/analytic) of the model. The results are presented in Sect. \ref{sec:results} and discussed in Sect. \ref{sec:discussion}, where we also describe the validity of our models with regards to all physical processes, which -- to our knowledge -- influence the results. We conclude in Sect \ref{sec:conclusions}.

\section{Model description}
\label{sec:model}
The model is built on two components: The first component consists of a protoplanetary disk model, including a single solid body embedded in the disk midplane. Its radial drift is calculated and its radially inward motion followed (Sect. \ref{ssec:radial_drift}). Once the temperature reaches a threshold value (set to 150K), the local disk temperature and radius of the body are used to calculate the body's thermal and compositional evolution during one timestep of the disk model. Two different modes for this second component are investigated: (a) a numerical model based on the cometary nucleus model by \citet{Marboeuf2012} (Sect. \ref{sec:cometary_nucleus}) or (b) an analytical expression treating the surface ablation (Sect. \ref{ssec:analytical_sublimation}). In both cases, mass loss from the body changes its radius, which in turn affects the drift speed.

\subsection{Disk}
\label{sec:disk_model}
Our $\alpha$-disk model assumes axis-symmetry, vertical hydrostatic equilibrium, flatness ($z\ll r$) and no self-gravity ($M_\text{Disk}\ll M_*$). The surface density $\Sigma\equiv \int_{-\infty}^\infty \rho(z) dz = \rho_0 H \sqrt{2\pi}$, where $\rho_0$ is the midplane density and $H$ is the vertical scale height, is evolved in time and the isothermal sound speed $c_s$ is frequently used to abbreviate the ideal gas law as $P=c_s^2 \rho$ \citep[see Sect. \ref{ssec:disk_evolution} or e.g.][for more details]{Armitage2019}.

We would like to point out that the purpose of the disk model in this paper is to simulate typical conditions and timescales in the disk midplane only, thus a simple model is sufficient. Of particular importance for this work is the thermal part of the disk model (see Sect. \ref{ssec:midplane_T}).

\subsubsection{Midplane temperature}
\label{ssec:midplane_T}
The midplane temperature $T$ is calculated, as in the model of \citet{Hueso2005}, by assuming that the disk is geometrically thin, heated viscously and by irradiation from the star, and is optically thick in the radial direction. Instead of solving the radiative transfer numerically, analytic expressions derived by \citet{Nakamoto1994} are used\footnote{Eq. 3.10 in \citet{Nakamoto1994}}. In their work, the mid-plane temperature is approximated as a sum of terms accounting for optically thick, optically thin, and stellar contributions:
\begin{equation}
\sigma T^4 = \frac{9}{8}\left(\frac{3}{8} \tau_R + \frac{1}{2\tau_p} \right) \Sigma \nu \Omega_\text{K}^2+\sigma T_l^4\,,
\label{eqn:midplane_temp}
\end{equation}
where $\sigma$ is the Stephan-Boltzmann constant, $\nu$ is the viscosity described in Sect. \ref{ssec:disk_evolution} and $\tau_P$ is the Planck mean optical depth which is assumed to be $\tau_P = 2.4 \tau_R$ as in \citet{Nakamoto1994}. $\tau_R$ is the Rosseland mean optical depth, which is defined in terms of the Rosseland mean opacity $\kappa_R$ as
\begin{equation}
\tau_R = \kappa_R \Sigma\,.
\end{equation} 
The Rosseland mean opacity is calculated using the modified Alexander/Cox/Stewart opacities by \citet{Bell1994}, 
\begin{equation}
\kappa_R = \kappa_i \rho_0^a T^b\,,
\end{equation}
where the exponents $a$ and $b$ and the factor $\kappa_i$ differ in different temperature regimes, i.e. depend on the gas state. The values for $a$, $b$ and $\kappa_i$ can be found in the appendix of \citet{Bell1994}.

$T_l$ is an effective temperature which includes effects of the irradiation by the star
\begin{equation}
T_l = T_*\left[\frac{2}{3\pi}\left(\frac{R_*}{r}\right)^3+\frac{1}{2}\left(\frac{R_*}{r}\right)^2\frac{H}{r}\left(\frac{d \ln H}{d \ln r} -1\right)\right]^{1/4}\,,
\label{eqn:tl_term}
\end{equation}
where the first term is due to irradiation onto the inner part of a flat disk by a finite sized star (with radius $R_*$) and the second term is accounting for irradiation onto the flared outer part. At all radii, we fixed $d \ln H / d \ln r = 9/7$ as in \cite{Hueso2005}. In contrast to their work, however, we did not include a molecular cloud temperature in equation \eqref{eqn:tl_term} and instead use a fixed floor value of \SI{10}{\kelvin} for the temperature.

\subsubsection{Disk evolution}
\label{ssec:disk_evolution}
For the disk, we assume an $\alpha$-viscosity $\nu = \alpha c_s H$, where $\alpha$ is a numerical factor on the order of \SI{e-4}{} to \SI{e-2}{} \citep{Shakura1973}. This viscosity, together with mass and angular momentum conservation, and approximating the orbital velocity of the gas to be Keplerian, lead to a diffusion equation for the disk \citep{Lynden-Bell1974,Pringle1981}
\begin{equation} 
\label{eq:disk_evol}
\frac{d \Sigma}{d t} = \frac{3}{r}\frac{\partial}{\partial r} \left( r^{1/2} \frac{\partial}{\partial r}\left(r^{1/2} \nu\Sigma\right)\right) + \dot\Sigma_w\,.
\end{equation}

We have added an external photo-evaporation term
\begin{equation}
\begin{cases}
 \dot\Sigma_w =0, &\text{if } r\leq r_g\\
\dot\Sigma_w \propto r^{-1}, & \text{otherwise}\,,
\end{cases}
\end{equation}
where the gravitational radius $r_g$ is taken to be equal to \SI{5}{\au} \citep{Veras2004}. The mass loss parameter $\dot M_\text{wind}$, corresponds to the mass that a disk extending to \SI{1000}{\au} would lose due to photo-evaporation. The actual mass loss due to photo-evaporation is approximately \SI{1}{\percent} of this value because the typical disk only extends to $\sim \SI{100}{\au}$. $\dot M_\text{wind}$ can be chosen to reproduce reasonable lifetimes \citep{Ribas2014}, which is the case for values $\sim \SI{e-7}{\umass\per \yr}$.

In our disk model, the disk evolution equation \eqref{eq:disk_evol} is solved numerically on a one dimensional, logarithmically spaced grid in radial direction \citep{Alibert2005, Alibert2013}.

\subsubsection{Radial drift}
\label{ssec:radial_drift}
Solid bodies in the disk feel a drag force caused by the different velocities of the gas ($v_g$) and the particles ($v_K$) which would move with Keplerian speed, in the absence of gas, whereas the former move with a slower velocity due to pressure support \citep{Weidenschilling1977, Whipple1972}.

To quantify this difference,
\begin{equation}
\eta \equiv 1- v_g/v_K \approx - \frac{r}{2 v_K^2\rho_0} \frac{dP}{dr}
\end{equation}
is defined, where the density $\rho_0$ and the pressure $P$ are the values taken at the location of the body.

The resulting radial drift depends on the drag regime. To discriminate between the different regimes, the radius $R$ of the solid body is compared to the mean free path $\lambda= m_{\text{H}_2}/(\pi d^2_{\text{H}_2} \rho_0)$. Here, $d_{\text{H}_2}$ and $m_{\text{H}_2}$ are the kinetic diameter and mass, respectively, of the hydrogen molecule, i.e. the dominant species in the disk. With that, the drag regime is determined by the following conditions:
\begin{itemize}
\item If $R<3\lambda/2$, we use the Epstein drag law
\item Else: \begin{itemize}
\item if $R_e <27$, where $R_e = 3 (v_K-v_g) R/(\lambda v_\text{therm})$ is the microscopic Reynolds number and $v_\text{therm}$ the midplane mean thermal velocity, the Stokes drag is used \citep{Rafikov2004},\footnote{This is an approximation introduced by \citet{Rafikov2004}, in the literature there is often an additional, intermediate drag regime used between the Quadratic and the Stokes regime \citep{Weidenschilling1977, Whipple1972}. Additionally, we use here the thermal velocity instead of approximating it as the sound speed. Furthermore, the definition of $R_e$ differs by a factor of two compared to \cite{Whipple1972}.}
\item if $R_e >27$, the bodies drag is governed by the quadratic drag law.
\end{itemize}
\end{itemize}

In this work, we use the radial drift formula that is used in \citet{Chambers2008} and is based on the solutions found by \citet{Adachi1976} and similarly by \citet{Nakagawa1986,Takeuchi2002}:
\begin{equation}
\label{eq:radial_drift}
\frac{dr}{dt} \approx \begin{dcases}
-\frac{2 r \eta}{t_\text{stop}}, &\text{(Quadratic regime)}\\
-\frac{2 r \eta}{t_\text{stop}} \left(\frac{s^2}{1+s^2}\right), &\text{(Epstein/Stokes regime)}\,,
\end{dcases}
\end{equation}
where
\begin{equation}
s= t_\text{stop} \Omega_\text{K}
\end{equation}
is the Stokes number. When switching from Stokes regime to the quadratic regime, the Stokes number should be large, such that there will be no discontinuity of the drift speed. This is the case in our application ($s \sim 5000$ for radii $\sim \SI{100}{\meter}$ for which the drag regime changes, see also Fig. \ref{fig:vdrift_regimes_irr_w1e-7}).

The stopping times differ in the three drag regimes and are given by \citep{Chambers2008, Takeuchi2002,Whipple1972} 
\begin{equation}
t_\text{stop} = \begin{dcases}
\frac{\rho_s R}{\rho_0 v_\text{therm}} & \text{(Epstein)}\\
\frac{2\rho_s R^2}{3\rho_0 \lambda v_\text{therm}}&\text{(Stokes)}\\
\frac{6\rho_s R}{\rho_0 (v_K - v_g)}&\text{(Quadratic)}\,,
\end{dcases}
\label{eq:stopping_time}
\end{equation}
where $R$ and $\rho_s$ are the radius and the density of the solid body.\footnote{As in \citet{Chambers2008}, the stopping time in the quadratic regime includes a factor of 6, in agreement with \citet{Whipple1972}, but in disagreement to the factor 5 in \citet{Rafikov2004}. Correspondingly the quadratic drag regime boundary is set to $R_e=27$ instead of $R_e=20$.} It can be easily verified that the drag regimes are chosen such that the stopping time is continuous.

The orbits of the bodies are assumed to be circular, thus the effect of eccentricity and inclination on the drift are omitted. This assumption is reasonable because the eccentricity and inclination get damped by gas drag on shorter timescales than the semi-major axis \citep{Adachi1976}.

We note, that the drift formula \eqref{eq:radial_drift} does not include the radial gas flow (e.g. \citet{Desch2017}), which does not have a large influence during the fast drifting phases and for large bodies (i.e. high Stokes numbers).

\subsection{Cometary nucleus model}
\label{sec:cometary_nucleus}
We describe here the cometary nucleus model \citep{Marboeuf2012} that we apply to solid bodies embedded in the protoplanetary disk. The model is able to include multiple volatile species, clathrates and amorphous water ice structures in a 1D \citep{Marboeuf2012} or 3D model \citep{Marboeuf2014c}. Here, we use the 1D model and pure crystalline water ice. Therefore, the equations can be simplified and, for completeness, are presented in that form here, with emphasis on the changes due to the presence of a disk.

The presence of the disk influences the energy sources available to the nucleus, which is discussed in Sect. \ref{ssec:energy_sources}. Then, the structure and physical model (Sect. \ref{ssec:structure}) are summarized. Finally, the possible formation of a dust mantle is described in Sect. \ref{ssec:dust_mantle_formation} and has a large impact on the resulting evolution of snowline crossing bodies.

\subsubsection{Energy sources}
\label{ssec:energy_sources}
In the comet related literature, there are usually three main sources of energy considered: solar radiation which heats the surface with energy propagating inward, internal heat release by radioactive isotopes contained in the cometary dust, and the release of latent heat originating from different possible phase changes (crystalline/amorphous ice, clathrates, sublimation) \citep{Klinger1981}.

When considering bodies smaller than \SI{100}{\meter}, radioactive heating is negligible: The simple estimates and values from \citet{Merk2003} \footnote{See their equation 2 with $KT\approx \SI{5}{\watt\per\meter}$, $\rho=\SI{0.5}{\gram\per\centi\meter\cubed}$, $t=0$} for heating by $^{26}$Al yield a net cooling for radii below \SI{1}{\km}, due to conduction and radiation from the surface. Because the heat produced by radioactive decay scales $\sim R^3$, whereas the cooling scales $\sim R^2$, radioactive heating becomes more important for larger objects with radii $> \SI{10}{\km}$.

The crystallization of amorphous water ice is dependent on the temperature and becomes efficient above \SI{100}{\kelvin} \citep{Schmitt1989}. It is not clear whether the initial structure of water ice inside comets is crystalline or amorphous. We assume an initial purely crystalline and clathrate-free water ice structure, or crystallization to have happened before the start of the calculation.

For a body embedded in the protoplanetary disk, additional energy sources are available: heat transfer (mainly by isotropic thermal radiation) from the gaseous disk and frictional heating. In contrast, direct irradiation from the sun is suppressed, as the disk is opaque in the midplane (see also Sect. \ref{ssec:structure} for implementation).

Frictional heating is caused by the different azimuthal velocities of the disk gas and the solid body. This process is negligible for the main part of this study and discussed in Sect. \ref{ssec:frictional_heating}.

Hence, the main source of energy for a small body relates to the thermal bath in which the body is moving. This justifies the use of a one dimensional model because of the invariance of heating on the orientation of the body in the disk. We use the local disk gas temperature as a surface temperature for our numerical model. This assumption is discussed in Sect. \ref{ssec:gas_v_surf_t}.

\subsubsection{Structure and physical model}
\label{ssec:structure}
\begin{figure}[ht]
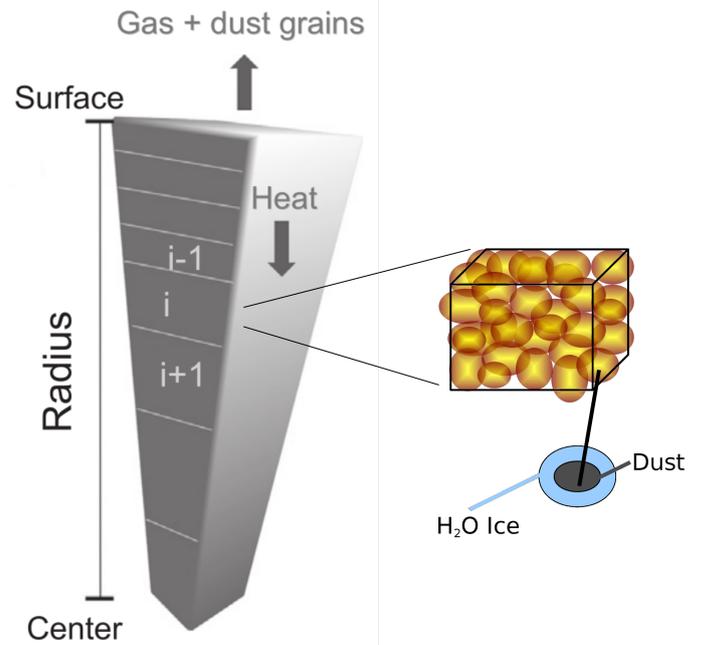

\center
\includegraphics[width=\linewidth]{{{radial_view_w_s_marboeuf_2014_no_co}.png}}
\caption{Schematic view of the structure model. Adapted from \citet{Marboeuf2014c}.}
\label{fig:structure_model}
\end{figure}

The body is composed of grains, consisting of refractory material, that are enclosed by a mantle of icy water following the model by \citet{Greenberg1988}. This structure is drawn schematically in Fig. \ref{fig:structure_model}.

For each layer, the diffusion of water vapor through the solid structure of grains is solved using the mass conservation equation. The only processes that can release or bind gas in our crystalline and clathrate-free model are water ice sublimation and condensation.

\begin{figure}[ht]
\center
\includegraphics[width=\linewidth]{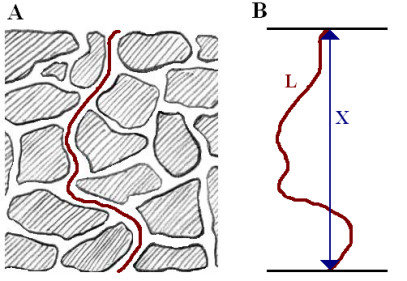}
\caption{Tortuosity of a path through a porous structure. In (A) a path through the material is shown, in (B) the length of the pore $L$ and the distance between the endpoints $X$ is indicated. Tortuosity is defined as $L/X$. Image adapted from \cite{OConnell2010} under a creative common licence (http://creativecommons.org/licenses/by/2.0).}
\label{fig:tortuosity}
\end{figure}

The flow of gas through the solid matrix can be in different flow regimes; free molecular (Knudsen) flow ($K^n>1$) or viscous flow ($K^n\ll1$), depending on the Knudsen number $K^n=\lambda / (2r_p)$, where $\lambda$ is the mean free path of the molecules and $r_p$ is the radius of a pore \citep{Knudsen1909}. In addition, we include a transition flow regime for Knudsen numbers $\SI{e-2}{}<K^n<1$ following \citet{Fanale1987}. One important quantity appearing in the expressions for the flow \citep{Marboeuf2012}, due to the influence of the structure of the pores, is tortuosity. Here, the arc-chord ratio definition of tortuosity is used, i.e. tortuosity is the ratio of the length of a pore to the distance between the endpoints (see Fig. \ref{fig:tortuosity}). 

Energy is conserved at each point inside the nucleus, where heat conduction is modeled using an empirical Hertz factor to account for porosity \citep{Davidsson2002,Prialnik2004a}. For detailed equations and explanations refer to \citet{Marboeuf2012}. 

In \citet{Marboeuf2012} the thermal boundary condition is given by a balance between the solar energy, sublimation of water ice at the surface (if no dust mantle is present), and thermal emission at the outermost layer of the nucleus. In the midplane of a gaseous disk, there is no direct irradiation, since the disk is opaque. Instead, the midplane temperature is used as a boundary condition at the surface of the nucleus and we do not solve the energy balance equation at the surface (this assumption is discussed in Sect. \ref{ssec:gas_v_surf_t}). The surface sublimation rate follows expression \eqref{eqn:analytical_subli}.

The gas pressure of the disk in the vicinity of the body is neglected, i.e. the partial pressure of water that is relevant for the sublimation rate is set to zero and the potential influence of the disk gas on diffusion of gaseous species in the interior is not considered. In our model the total amount of gas in the interior is given by the tracked gas flow of the cometary nucleus model and is not including disk gas. We discuss the impact of disk material on the dominating surface sublimation rate in Sect. \ref{ssec:water_vapor_pressure}.

\subsubsection{Dust mantle formation}
\label{ssec:dust_mantle_formation}
The solid dust grains that are freed from the rigid structure by sublimation of water ice in the interior can either be ejected from the nucleus or they can accumulate at the surface. The mechanisms for this accumulation are reviewed in \citet{Prialnik2004a}. To summarize, there are multiple drivers of dust mantle formation that simultaneously appear in a body that undergoes sublimation.

Firstly, gas drag pulls the freed grains outward, but gravity counteracts this process. The magnitude of the gas drag force depends on the grain size, hence there is a critical radius $r_c$ of grains that can be ejected. For a slow spinning nucleus the centrifugal force can be neglected, thus \citep[equation 7]{Rickman1990}
\begin{equation}
r_c = \frac{3}{8} \frac{C_{D\text{,Kn}} m_{\text{H}_2\text{O}} \phi_{\text{H}_2\text{O}} v_{\text{H}_2\text{O}}}{\rho_\text{grain} G \frac{M_\text{nucleus}}{R_\text{nucleus}^2}}\,,
\end{equation}
where $C_{D\text{,Kn}}\sim 2$ is the drag coefficient in the free molecular (Knudsen) flow regime \citep{Prialnik2004a}, $R_\text{nucleus}$ and $M_\text{nucleus}$ are the radius and the mass of the whole nucleus, $m_{\text{H}_2\text{O}}$ the molar mass, $\phi_{\text{H}_2\text{O}}$ the molar flow, and $v_{\text{H}_2\text{O}}$ the velocity of water vapor.\footnote{This expression differs from the one in \cite{Rickman1990}, since, in our case, the gas flow is numerically modeled throughout the nucleus and can be used directly instead of analytically estimating it.} Grains with radii larger than $r_c$ do not get ejected but instead settle on the nucleus' surface, already depleted of ice by sublimation. Hence, a porous dust mantle forms on the surface. \cite{Huebner2006} remark that $r_c$ only gives an upper size limit, for escaping grains, but smaller grains are not necessarily escaping, as the flow of gas thins above the surface and the grain might fall back onto the nucleus. Furthermore, already in early studies investigating this process, e.g. \citet{Brin1979}, it was noted that for large dust-to-volatile mass ratio it is impossible to blow off all the freed dust, even though the particles might have radii smaller than $r_c$.

The second process comes into play if a dust mantle already exists. The accumulated grains on the surface will interfere with the liberated grains, such that they can no longer pass through the less porous mantle. Hence, they are trapped within the nucleus and further increase the size of the mantle \citep{Shulman1972,Rickman1990}.

Furthermore, the dust mantle can break under the gas flow, or its cohesive strength can be large enough to trap not only the dust, but the gas as well \citep{Huebner2006}. In our model, no cohesive forces between the grains are taken into account \citep{Marboeuf2012}. Therefore, we test in section \ref{ssec:dust_mantle_influence} three cases: (a) the nominal case for which no initial dust mantle is present nor is it allowed to form subsequently, i.e. all the freed dust is lost, (b) an unstable dust mantle case, for which no cohesion forces are taken into account but particles larger than $r_c$ are assumed to fall back onto the surface after ejection thereby forming a dust mantle over time, and (c) a constant dust mantle case, with a fixed thickness over the full evolution of the body. In case (c) most of the ejected dust is still lost, but a fraction is kept to keep the artificial constant mantle thickness. These cases differ compared to the work of \cite{Schorghofer2008} who assumed that no dust is lost. This is essentially related to the size of the bodies considered. Small bodies with sizes below hundreds of meters undergoing sublimation (considered in this work) can lose their dust, but larger bodies (considered in \cite{Schorghofer2008}) will be able to keep their dust due to the increased gravity (i.e. $r_c \propto R_\text{nucleus}^2/M_\text{nucleus}$ is smaller than all the typical grain sizes).

\subsection{Analytical surface ablation model}
\label{ssec:analytical_sublimation}
Instead of invoking the full model from \citet{Marboeuf2012} which is described in Sect. (\ref{sec:cometary_nucleus}), an analytic model for the sublimation of water ice from the surface, i.e. ablation, is outlined here, which can be tested against similar models from the literature, e.g. \citet{DAngelo2015}, or our full model that includes the very same surface sublimation term.

For a body without a mantle, ablation follows the kinetic theory expression, also known as the Hertz-Knudsen-Langmuir formula, for a free sublimation rate \citep[e.g.][]{Hertz1882,Delsemme1971,Schorghofer2008,Marboeuf2012}
\begin{equation}
\varphi(T)=\frac{P^s(T)}{\sqrt{2\pi m_{\text{H}_2\text{O}}R_gT}} \quad\left(\SI{}{\mole\per\meter\squared\per\second}\right)\,,
\label{eqn:analytical_subli}
\end{equation}
where $P^s$ is the water vapor sublimation pressure (\SI{}{\pascal}), $m_{\text{H}_2\text{O}}$ is the molar weight of water and $R_g$ is the ideal gas constant (\SI{}{\joule\per\mole\per\kelvin}). Equation \eqref{eqn:analytical_subli} is valid assuming zero partial pressure of water in vicinity of the body. We discuss this approximation in Sect. \ref{ssec:water_vapor_pressure}. For non-zero pressure with the same temperature, the difference between pressures replaces $P^s$ in the equation.

If this amount of water is removed from a layer with thickness $\delta \ll R$ at the surface, the total water mass loss is
\begin{equation}
\frac{dm}{dt}\bigg\vert_{\text{H}_2\text{O}} = \varphi(T)\, m_{\text{H}_2\text{O}} \, 4\pi R^2\,.
\end{equation}
For the refractory part (i.e. dust) of the structure, we assume that the grains are freed in the surface layer and matter gets released immediately adding their contribution to the total mass loss. This can be compared to the case without dust mantle formation of the cometary nucleus model (see Sect. \ref{ssec:dust_mantle_formation}).

Expressed as a decrease in radius, we can write
\begin{equation}
\frac{dR}{dt} = \varphi(T)\, \frac{m_{\text{H}_2\text{O}}}{\rho_{\text{H}_2\text{O}\text{,matrix}}}\,,
\label{eqn:analytical_drdt}
\end{equation}
where $\rho_{\text{H}_2\text{O}\text{,matrix}}$ is the macroscopic water density in the matrix (taking into account porosity). The initial conditions shown in table \ref{tab:nucleus_initial} yield a $\rho_{\text{H}_2\text{O}\text{,matrix}} = \SI{276}{\kg\per\cubic\meter}$. At fixed porosity, increasing the amount of refractory components reduces the available water, hence reducing $\rho_{\text{H}_2\text{O}\text{,matrix}}$ and increasing the total mass loss. The assumption that the dust is freed with the sublimation of the ice is not valid for high dust to water ratios, since the cohesive forces between dust particles would become relevant.

\begin{figure}[ht]
\center
\includegraphics[width=\linewidth]{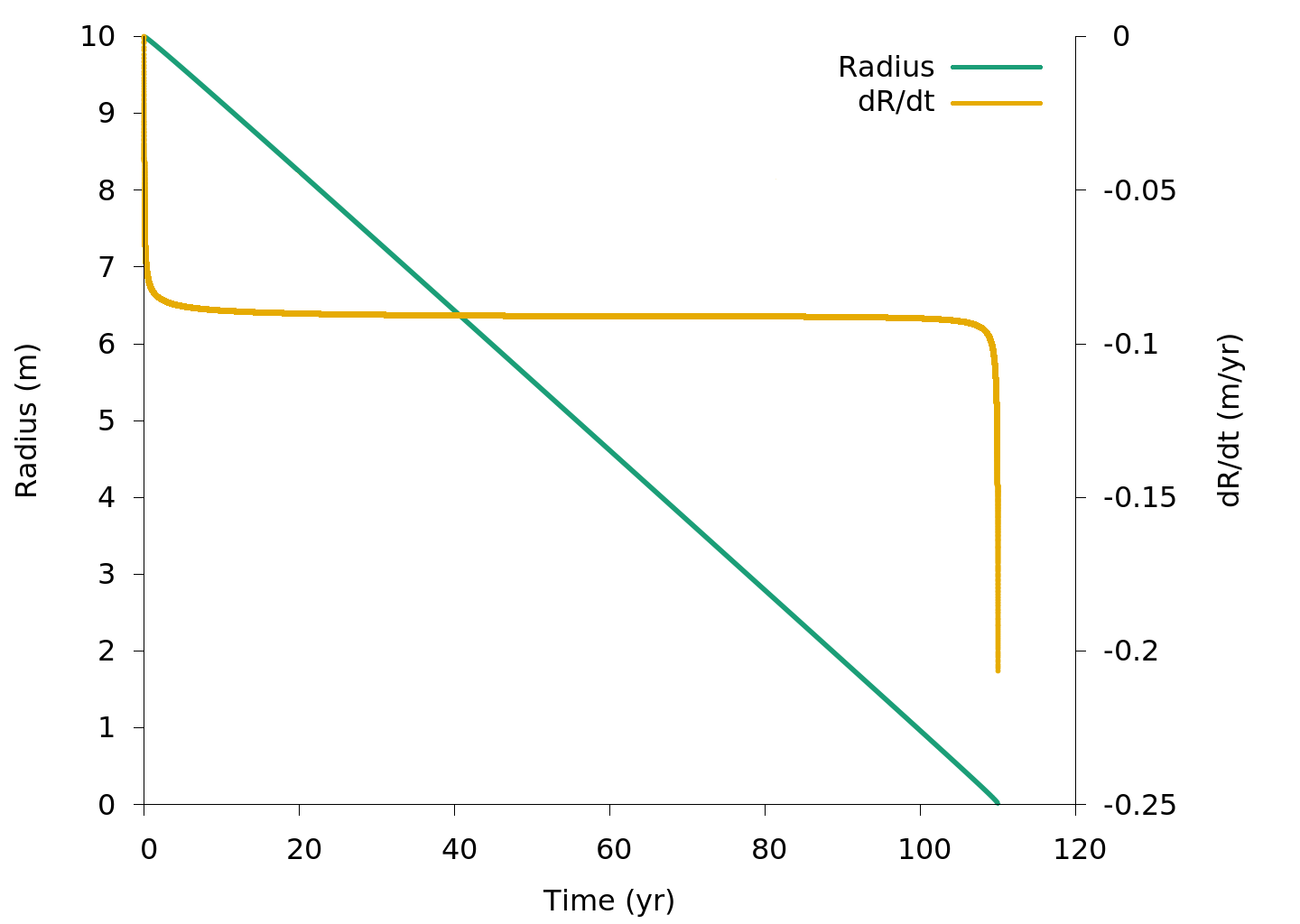}
\caption[Almost linear decrease in radius over time (green line, left axis) for a fixed surface temperature of \SI{169}{\kelvin} using the cometary nucleus model. The derivative $dR/dt$ is plotted in orange (right axis).]{Almost linear decrease in radius over time (green line, left axis) for a fixed surface temperature of \SI{169}{\kelvin} using the cometary nucleus model. The derivative $dR/dt$ is plotted in orange (right axis).}
\label{fig:t_R_dRdt}
\end{figure}

It is noteworthy that if the temperature is kept constant and the body has a homogeneous structure, expression \eqref{eqn:analytical_drdt} is independent of the body's total radius, leading to a constant decrease in radius over time. Fig. \ref{fig:t_R_dRdt} shows the result of a cometary nucleus model run of an initially \SI{10}{\meter} sized body, which exhibits this behavior and motivates the analytic sublimation formula. This is true, as long as (i) the total radius is much larger than the radial extent of the surface layer ($R\gg \delta$) and (ii) there are no interior temperature gradients that would change the surface temperature.

Furthermore, we would like to point out that the analytic surface sublimation model is identical to the full cometary nucleus model (Sect. \ref{sec:cometary_nucleus}) assuming: (i) no heat transport of any sort, i.e. the temperature inside the body's structure is the same as in the disk, (ii) no other species than pure crystalline water ice and dust to be present and (iii) no mantle at the surface to form.

\subsection{Initial conditions}
\label{ssec:initial_conditions}

\subsubsection{Disk}
The initial gas surface density of the disk is given by a power law with exponential outer cut-off boundary \citep[as proposed by][]{Andrews2010} and a normalization constant $\Sigma_0$, corresponding to the surface density at approximately \SI{5.2}{\au}, which determines the total disk mass.
\begin{equation}
\Sigma(r) = \Sigma_0 \left(\frac{r}{\SI{5.2}{\au}}\right)^{-\beta} \exp{\left[- \left(\frac{r}{R_\text{out}}\right)^{(2-\beta)}\right]}\,,
\end{equation}
where $R_\text{out}$ is a constant exponential cut-off radius, $\beta$ is the power law exponent, determining the slope of the surface density profile. The disk evolution is then given by the Shakura-Sunyaev $\alpha$ parameter and the photo-evaporation (Sect. \ref{ssec:disk_evolution}). We leave $\alpha$ fixed and run simulations with (nominal) and without photo-evaporation. The values, which are fixed in all results in this paper, can be seen in table \ref{tab:disk_initial}. The initial total gas mass in the disk is accordingly \SI{0.05}{\umass}, which is the disk mass that \cite{Weidenschilling1977} uses for the minimum mass solar nebula (MMSN). The star was assumed not to evolve during the disk's lifetime and the temperature and radius values are taken at a time of \SI{1}{\mega\yr} of stellar evolution according to \cite{Baraffe2015}.

In order to gauge the influence of the initial parameters, we varied the total mass, lifetime and heating mode of the disk and the results and changes to the nominal parameters can be found in Sect. \ref{ssec:disk_influence}.

\newcommand\T{\rule{0pt}{2.6ex}}       
\newcommand\B{\rule[-1.2ex]{0pt}{0pt}} 
\begin{table}
\center
\caption{Physical parameters for the nominal disk initial structure and evolution}
\label{tab:disk_initial}
\begin{tabular}{l  r}
\hline
\hline
Parameter & Value \T\B \\
\hline
Stellar mass & \SI{1}{\umass} \T\\
Stellar radius & \SI{2.416}{R_\odot}$^{(a)}$\\
Stellar effective temperature & \SI{4377}{\kelvin}$^{(a)}$ \\
Helium fraction & \SI{0.24}{} \\
Power law slope $\beta$ & \SI{0.9}{}$^{(b)}$ \\
Cut-off radius $R_\text{out}$ & \SI{50}{\au}$^{(b)}$\\
Inner boundary radius & \SI{0.03}{\au} \\
Surface density at \SI{5.2}{\au} $\Sigma_0$ & \SI{268.5}{\gram\per\centi\meter\squared}$^{(c)}$\\
Shakura-Sunyaev $\alpha$-viscosity & \SI{2e-3}{} \\
Photo-evaporation parameter $\dot{M}_\text{wind}$ & \SI{e-7}{\umass\per \yr} \B\\
\hline
\T
\textit{References.}\\
$^{(a)}$ \cite{Baraffe2015} at \SI{1}{\mega\yr}; & \\
$^{(b)}$ \cite{Andrews2010}; & \\
$^{(c)}$ MMSN \citep[e.g.][]{Weidenschilling1977} & \B\\
\hline
\end{tabular}
\end{table}

\subsubsection{Solid body}
To reduce complexity, we chose to model a body consisting only of water ice and dust, without any other volatile species. Water is the main volatile component \citep[see][about the composition of planetesimals in disks]{Marboeuf2014} and the last one to sublimate. Using the parameters listed in table \ref{tab:nucleus_initial}, the resulting total density of the body is $\sim \SI{0.42}{\g\per\centi\meter\cubed}$ which is of the same order of magnitude as the recently found bulk density of \SI{0.533(6)}{\gram\per\cm\cubed} \citep{Patzold2016} and the previous value of \SI{0.470(45)}{\gram\per\cm\cubed} by \citet{Sierks2015} of the comet 67P/Churyumov-Gerasimenko. Addition of other volatile species would increase the density to values even closer to these measurements. We chose to represent realistic dust to ice mass ratios ($\sim 1$) \citep{Marboeuf2014} instead of tuning the ratio to represent measured bulk densities.

The initial location of the body in the disk is set to a distance \SI{10}{\percent} further away from the star than the snowline, unless otherwise stated. This starting position allows the body to relax to the environment so that initial conditions are forgotten by the time we start computing evaporation.

For heat capacities and conductivities of dust and water ice, we adopted the values listed in \citet{Marboeuf2012} and references therein.

\begin{table}
\center
\caption{Physical parameters of the cometary nucleus}
\label{tab:nucleus_initial}
\begin{tabular}{l  c}
\hline
\hline
Parameter & Value \T\B \\
\hline
Initial nucleus porosity & \SI{70}{\percent}\T\\
Dust mantle porosity & \SI{70}{\percent}\\
Tortuosity&$\sqrt{2}^{(a)}$\\
Initial dust/ice mass ratio&\SI{1}{}$^{(b)}$\\
Water ice bulk density& \SI{920}{\kg\per\meter\cubed}$^{(c)}$\\
Dust bulk density&\SI{3000}{\kg\per\meter\cubed}$^{(c)}$\\
Enthalpy of sublimation&$\left[51983.9-20.0904\, T\right]$  \SI{}{\joule\per\mole}$^{(d)}$\\
Heat conductivity&$\left[0.0028 + 1.3/T \right]$ \SI{}{\watt\per\meter\per\kelvin}$^{(c)}$\\
Volumetric heat capacity&$\left[ 1582\, (114.8 + T)\right]$ \SI{}{\joule\per\kelvin\per\meter\cubed}$^{(c)}$\B\\
\hline
\T
\textit{References.}\\
\multicolumn{2}{l}{$^{(a)}$ \cite{Carman1956,Mekler1990,Kossacki2006};}\\
$^{(b)}$ \cite{Marboeuf2014};&\\
$^{(c)}$ \cite{Marboeuf2012};&\\
\multicolumn{2}{l}{$^{(d)}$ \cite{Washburn1928,Delsemme1971}}\B\\
\hline
\end{tabular}
\end{table}

\section{Results}
\label{sec:results}
We first (Sect. \ref{ssec:model_comparision}) present the test cases comparing the two different sublimation models described in Sects. \ref{ssec:analytical_sublimation} and \ref{sec:cometary_nucleus}. Then, we study which bodies are able to cross the snowline (Sect. \ref{sec:snowline_crossing}). Finally, the results of simulated bodies crossing the snowline that were mainly obtained with the full cometary nucleus model for different varied quantities are presented in Sect. \ref{sec:snowline_crossing}.

\subsection{Comparision between the two sublimation models}
\label{ssec:model_comparision}

\begin{figure}[ht]
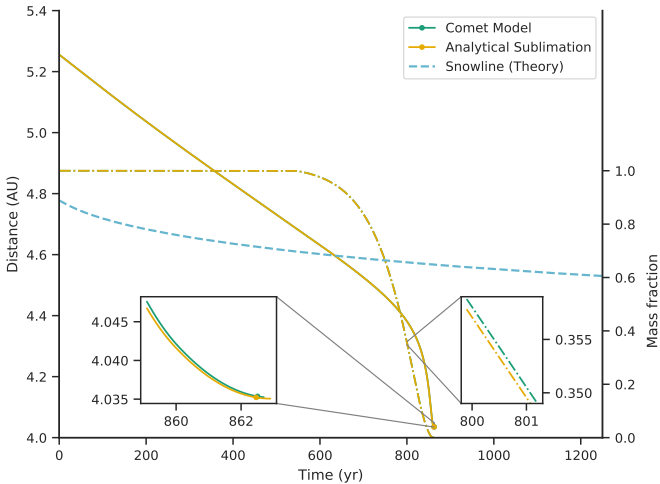

\center
\includegraphics[width=\linewidth]{{{fig_4_compare_analytical_10m_nominal_disk}.pdf}}
\caption{Comparison of the comet model solving the internal structure and the analytical solution (equation \ref{eqn:analytical_subli}) for \SI{10}{\meter} sized bodies. The solid lines show the distance to the star (left axis) with dots representing the locations where the bodies shrank to a size of \SI{10}{\centi\meter} while the dash-dotted lines show the remaining mass fraction (right axis). The lines of the two different model solutions are essentially indistinguishable. The initial position is \SI{10}{\percent} above the snowline location at time zero in the nominal disk. The barely visible kink in the mass fractions at \SI{600}{\yr} is due to reaching the threshold temperature of \SI{150}{\kelvin}, where the sublimation models are started.}
\label{fig:comparision_analytical_10m}
\end{figure}

Fig. \ref{fig:comparision_analytical_10m} shows the results for a test case, in which we placed a body with an initial radius of \SI{10}{\meter} and the composition shown in table \ref{tab:nucleus_initial} into the nominal disk (see table \ref{tab:disk_initial}). The initial semi-major axis is set to \SI{6}{\au} at time zero of the disk evolution. We find almost indistinguishable outcomes between the analytical surface sublimation model and the cometary nucleus model for this particular test.

\begin{figure}[ht]
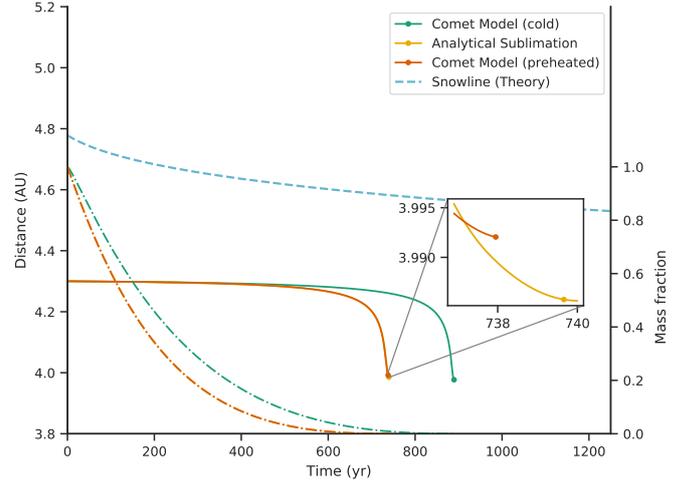

\center
\includegraphics[width=\linewidth]{{{fig_5_compare_analytical_100m_nominal_disk}.pdf}}
\caption{As Fig. \ref{fig:comparision_analytical_10m}, but for a \SI{100}{\meter} sized body. The initial position is chosen starwards of the snowline, i.e. at \SI{4.3}{\au}. One of the cometary nucleus model runs is started with a low initial bulk temperature of \SI{20}{\kelvin} (green line), whereas the other is pre-heated to the local gas temperature at the starting location (\SI{170.16}{\kelvin}), which is the implicit assumption of the Analytical Sublimation. The lines of the analytical sublimation model and the preheated comet model are barely distinguishable. The local gas temperatures at the end of the calculation are \SIlist[list-units=single,list-final-separator = {, }]{174.95;175.07;174.94}{\kelvin} for the preheated, analytical and the cold model respectively.}
\label{fig:comparision_analytical_100m}
\end{figure}

For the larger, \SI{100}{\meter} radius case (Fig. \ref{fig:comparision_analytical_100m}), the drift timescale is much larger. To save computation time, the body is initially positioned closer to the star than the initial snowline, namely at \SI{4.3}{\au}. The initial bulk temperature for the analytical sublimation is, by construction, assumed to be equal to the local disk gas temperature. To estimate the influence of the initial temperature of the body, we ran two different cases with the cometary nucleus model: one with a pre-heated body, i.e. the initial bulk temperature is set to \SI{170}{\kelvin}, which is the local gas temperature, and one without pre-heating, i.e. an initial bulk temperature of \SI{20}{\kelvin}. The results are discussed in Sect. \ref{ssec:internal_evolution}.

For equal initial conditions and under the assumption of initial homogeneous temperature in the nucleus, the analytical solution for the sublimation, i.e. equation \eqref{eqn:analytical_subli}, and the cometary nucleus model do agree well in the tested size range of bodies (i.e. meters to \SI{100}{\meter}). The agreement worsens with increasing size even in the absence of a dust mantle (see Sect. \ref{ssec:dust_mantle_influence}).

\subsection{Snowline versus drift velocity}
\label{sec:viceline_vdrift}
\begin{figure}[ht]
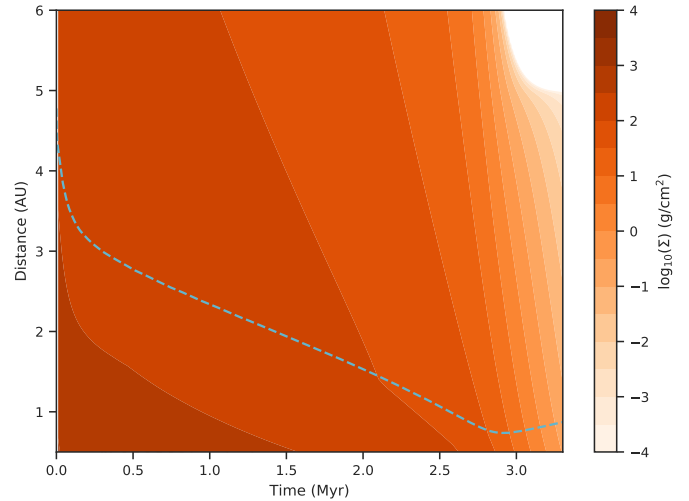

\includegraphics[width=\linewidth]{{{fig_6_sigma_nominal_disk}.pdf}}
\caption{Surface density evolution for the nominal disk. The dashed, blue line shows the snowline position.}
\label{fig:results_disks}
\end{figure}

Temperature and pressure determine the classical snowline position during the evolution of the disk. In our nominal disk (see table \ref{tab:disk_initial}), the classical water ice line, i.e. the snowline, was determined (see Fig. \ref{fig:results_disks}). Due to external photo-evaporation, the disk vanishes almost completely after \SI{2.8}{\mega\yr}. As the inner disk surface density decreases, direct irradiation from the central star can invert the cooling of the disk to a heating (via the direct irradiation included in $T_l$ in equation \ref{eqn:midplane_temp}) in the inner region. Thus, the snowline motion reverts as well. This would not happen in a disk without photo-evaporation, where the disk gradually thins out as a result of the viscous evolution only, i.e. depending solely on $\alpha$, and the snowline motion never changes direction.

\begin{figure}[ht]
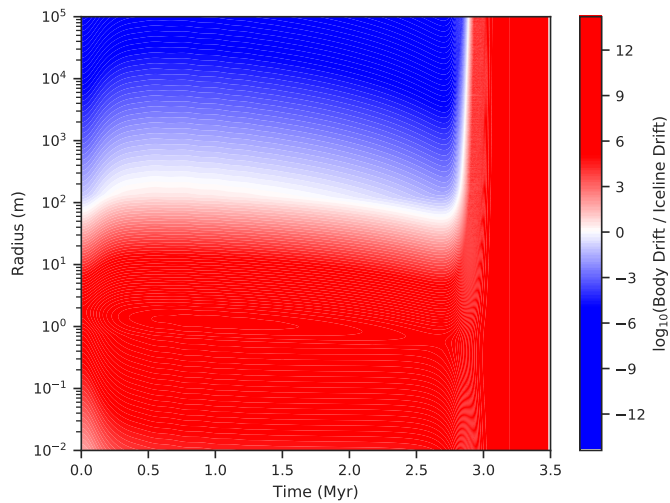
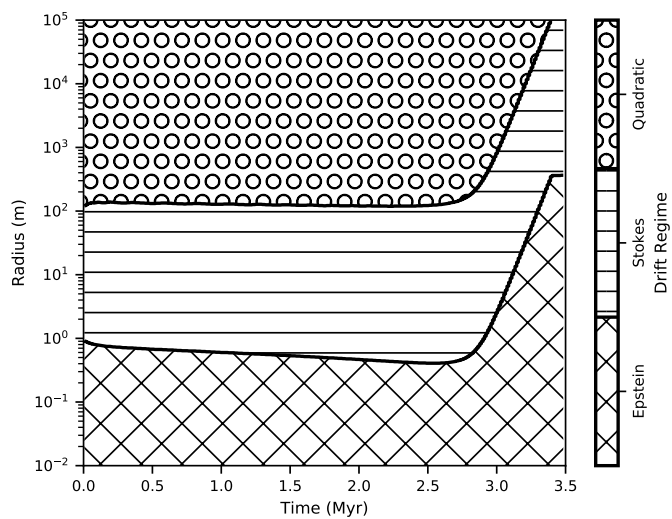

\subfigure[Drift velocity compared to water snowline velocity\label{fig:vdrift_irr_w1e-7}]{\makebox[\linewidth]{
\includegraphics[width=\linewidth]{{{fig_7a_results_vdrift_Nominal}.pdf}}
}}
\subfigure[Drag regime\label{fig:vdrift_regimes_irr_w1e-7}]{\makebox[\linewidth]{
\includegraphics[width=\linewidth]{{{fig_7b_results_vdrift_regimes_Nominal}.pdf}}
}}
\caption{Drift velocity and regime in an irradiated disk with photo-evaporation. All the bodies with sizes in the red area in Fig. (a) cross the snowline, since they drift faster than it moves towards the central star. The snowline is determined using tabulated values for the temperature and pressure. After approximately \SI{2.4}{\mega yr}, the snowline position starts to move away from the star, due to the disks dispersal. Thus, the ratio of the body's drift speed to the snowline speed is negative and the log in Fig. (a) is no longer defined and the ratio is set to a value of 12 to indicate that all the bodies will cross the snowline in that phase. In order to smooth out numerical artifacts, we applied a Gaussian filter in horizontal direction.}
\label{fig:vdrift_w1e-7}
\end{figure}

In the nominal disk model we calculated the drift speed (see Sect. \ref{ssec:radial_drift}) of solid bodies in the size range from \SI{e-2}{\meter} to \SI{e5}{\meter} over time, as well as the change of the snowline position. For objects bigger than \SI{e5}{\meter}, gas drag is not the relevant source of migration, but the torque exerted by density waves (type I migration) \citep{Goldreich1979, Ward1997}. The ratio of the body's drift speed to the snowline speed is shown in Fig. \ref{fig:vdrift_irr_w1e-7}. Important for our goal is the size range where the transition from bodies moving slower than the snowline to faster than snowline speed lies. In Fig. \ref{fig:vdrift_irr_w1e-7} the color code is chosen such that this transition lies in the white region. We found that planetesimals with $R \gtrsim \SI{100}{\meter}$ will no longer drift towards the star fast enough to cross the snowline, thus the water ice on these bodies will never sublimate. To help interpret the figure, the drag regime of the different sized bodies is plotted in Fig. \ref{fig:vdrift_regimes_irr_w1e-7}. A size of roughly \SI{100}{\meter} happens to coincide with the transition from Stokes to quadratic drag regime emphasizing the need to take into account the different drag regimes.

\subsection{Parameter study of snowline crossing bodies}
\label{sec:snowline_crossing}
In this part of the results section we present the evolution of drifting solid bodies in the protoplanetary disk. These results -- obtained using the cometary nucleus model -- are presented in Sect. \ref{ssec:initial_radius_dependence}, \ref{ssec:disk_influence} and \ref{ssec:dust_mantle_influence}, in which the radius, disk conditions and dust mantle properties are varied.

\subsubsection{Initial radius dependence}
\label{ssec:initial_radius_dependence}
\begin{figure}[ht]
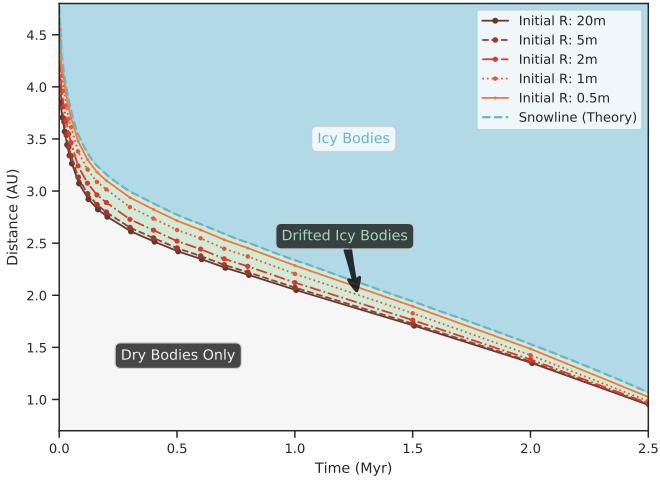
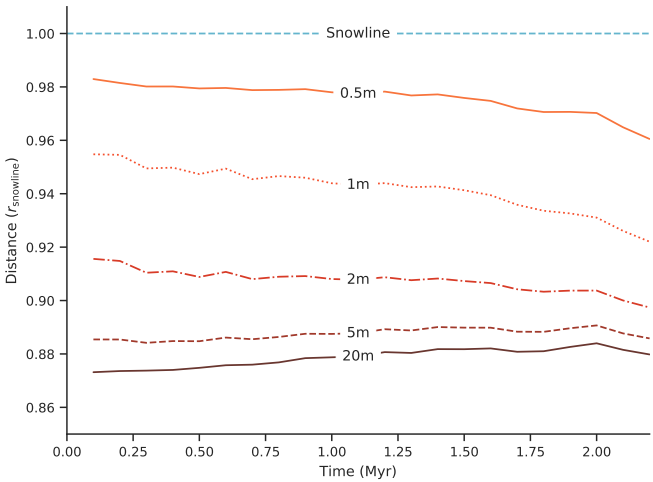

\center
\subfigure[Locations of complete disintegration for different initial sizes\label{fig:results_radii_au}]{\makebox[\linewidth]{
\includegraphics[width=\linewidth]{{{fig_8a_results_R0.5to20_nominal_disk}.pdf}}
}}
\subfigure[As panel (a), but with distance measured relative to the snowline position \label{fig:results_radii_rel}]{\makebox[\linewidth]{
\includegraphics[width=\linewidth]{{{fig_8b_results_R0.5to20_nominal_disk_rel}.pdf}}
}}
\caption{Comparison of locations of complete disintegration of different sized bodies without a dust mantle. In panel (a) the distance to the star is measured in \SI{}{\au} and the dots represent the locations where the body shrank to a size of \SI{10}{\centi\meter}. The dashed, cyan line indicates the evolving position of the $P$-$T$ tabulated snowline, and the regions where only icy solid bodies, only water depleted solid bodies, and the region that is injected with drifting icy bodies are colored and labelled. In panel (b), the same data is shown but measured in units of the evolving, tabulated snowline position (\SI{1}{} corresponds to the snowline position, \SI{0}{} to the central star).}
\label{fig:results_radii}
\end{figure}

Figs. \ref{fig:results_radii_au} and \ref{fig:results_radii_rel} show the innermost locations reached by different sized bodies, drifting from outer regions of the disk. In the following, we call this position the \textit{location of complete disintegration}. To get the results, the full cometary nucleus model mode was used. The composition of the different sized bodies was assumed to be equivalent and corresponds to the values given in Sect. \ref{ssec:initial_conditions} and table \ref{tab:nucleus_initial}. No dust mantle is present in all shown cases, i.e. dust mantle formation is excluded.

When the body reaches high enough temperatures it undergoes ablation and thus loses mass. After shrinking to a radius of \SI{10}{\centi\meter} the location is marked as a dot in Fig. \ref{fig:results_radii_au}. This location is considered to be the the location of complete disintegration, since a centimeter sized icy body at those temperatures and pressures has a very short lifetime \citep[e.g.][]{Lichtenegger1991}. \SI{20} bodies are modeled starting at  different times over the disk lifetime for each evaluated size. Initially, the bodies are located \SI{10}{\percent} further away from the star than the snowline location at the specific starting time.

The lowest included initial radius is \SI{0.5}{\meter}. Due to numerical and physical assumptions of the model, such as not tracking single grains, lower initial radii are excluded and these pebble sized objects are the main subject of other studies \citep[e.g.][]{Drazkowska2017, Schoonenberg2017}.

It can be seen in Fig. \ref{fig:vdrift_irr_w1e-7} that bodies with radii on the order of one meter drift the fastest in the protoplanetary disk \citep[see also][]{Weidenschilling1977,Adachi1976}. Bodies with radii lower than one meter drift slower and thus only cover a small distance after crossing the snowline. The smallest body in our dataset, with an initial radius of \SI{0.5}{\meter}, loses all of its mass and stops very close to the snowline due to its relatively slow drift. The tabulated snowline position values and the sublimation are calculated independently. Therefore, the good agreement of the snowline location in Fig. \ref{fig:results_radii} with the location of complete disintegration of the \SI{0.5}{\meter} sized body shows that the tabulated snowline position is a reasonable choice of reference.
	
A larger than meter-sized body with identical composition will also undergo sublimation and thus lose mass. As a consequence, it gradually speeds up until it reaches maximal drift speed at a radius of one meter. This size will be reached closer to the star than the equilibrium position of the snowline. Thus, the object will ultimately drift farther than an initially smaller body until complete disintegration. The difference in the position of complete disintegration will decrease with increasing size, because the initial drift slows down and thus only little distance is traveled before reaching a smaller size. However, the difference will stay bigger than zero, and therefore bigger bodies always cross a larger distance before they completely disintegrate. This asymptotic behavior can be seen in Figs. \ref{fig:results_radii} and \ref{fig:results_radii_rel}. Since the difference of the position of disintegration between a five meter sized and a ten meter sized body is negligible compared to other effects (see e.g. Sect. \ref{ssec:dust_mantle_influence}) and due to the numerical cost of simulating a large body, no larger sizes were included. There is no reason to expect the position of disintegration of larger bodies to change significantly compared to the one of ten meter sized bodies up to the \SI{100}{\meter} size boundary, where the bodies can no longer cross the snowline by radial drift (see Sect. \ref{sec:viceline_vdrift}).

To show that the results can be well decoupled from the disk evolution, Fig. \ref{fig:results_radii_rel} shows the same results as Fig. \ref{fig:results_radii_au}, but instead of measuring the distance from the central star in units of \SI{}{\au}, it is measured in units of the snowline position $r_\text{snowline}$ (\SI{1}{} corresponds to the snowline location, \SI{0}{} to the central star). The ratio of the location of complete disintegration to the snowline position stays approximately constant in time.

We would like to point out that this behavior is only found if the bodies are not covered by dust mantles. For bodies with dust mantles, a similar size-dependent behavior is only recovered for exactly equal mantle thicknesses using the constant dust mantle mode described in Sect. \ref{ssec:dust_mantle_influence}. However, the scaling of the mantle thickness depending on size would be the dominant factor but is to our knowledge not well constrained.

\begin{figure*}[ht]
\center
\includegraphics[width=\linewidth]{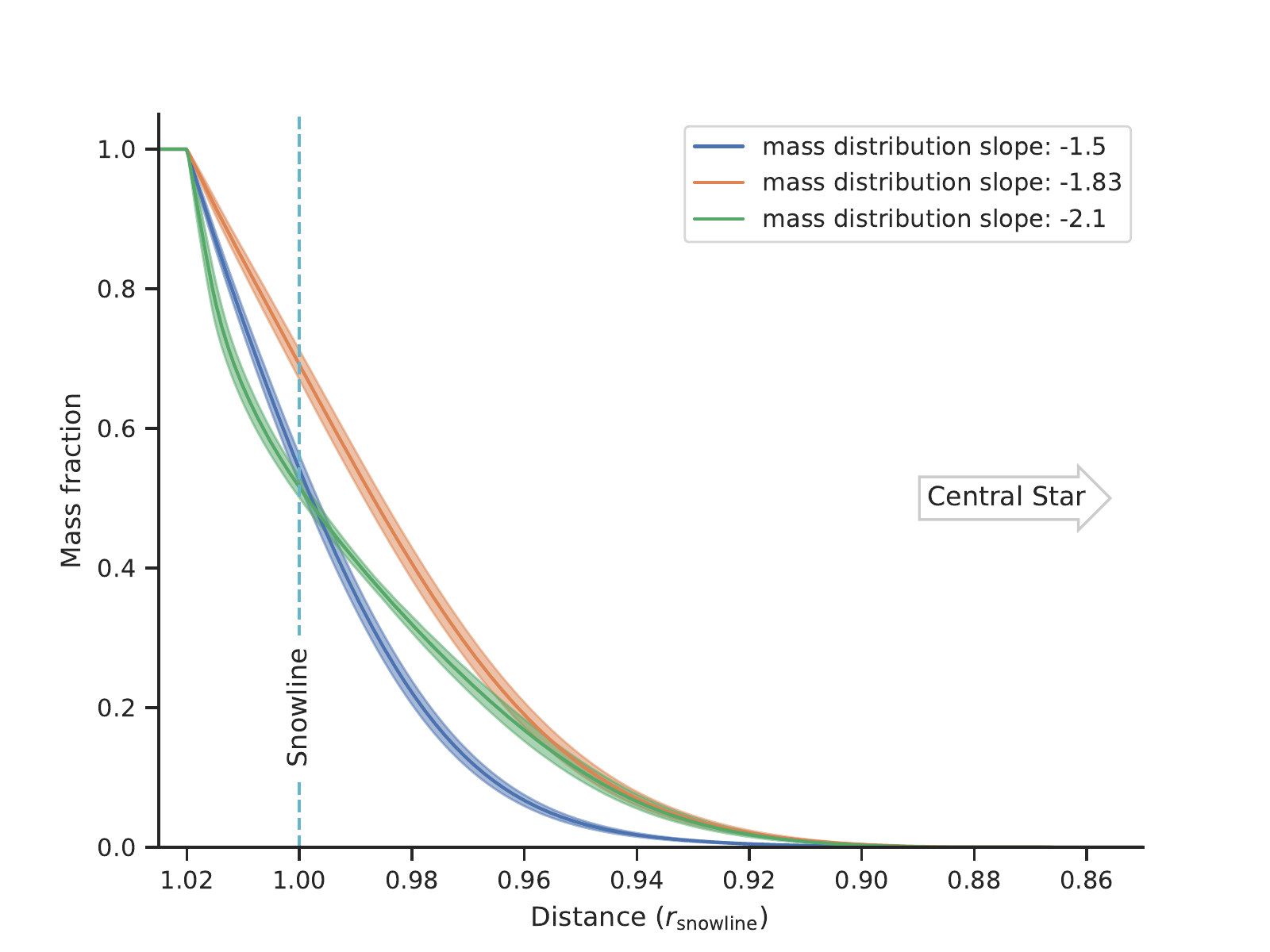}
\caption{Remaining mass fraction in the overall population of bodies crossing the snowline ($\SI{1}{\kilogram} \le m\le \SI{1e9}{\kilogram}$) with shaded bands indicating the standard deviation due to the evolving disk. The mass shown is an integral over a distribution of masses with the indicated power-law slope and a mean over time in the disk. More details can be found at the end of Sect. \ref{ssec:initial_radius_dependence}.}
\label{fig:results_mass_deposition_slope}
\end{figure*}

The aforementioned time-decoupling of the effect by using units of snowline distance can be used to tentatively explore the overall mass fraction of drifting bodies that should be found at a given distance from the star -- measured in units of the snowline position -- at all times in the disk (Fig. \ref{fig:results_mass_deposition_slope}). The mass fraction value shown is an integral over the assumed distribution of bodies (see Sect. \ref{ssec:collisions}), which was cut such that all included sizes do cross the snowline at all times of the nominal disk evolution (i.e. \SI{1}{\kilogram} to \SI{1e9}{\kilogram} corresponding to \SI{8}{\centi\meter} to \SI{76}{\meter}). For simplicity, the density was fixed to the nominal value of \SI{0.422}{\gram\per\centi\meter\cubed} and the analytical sublimation model was used. To help interpret the results, we note that the largest bodies which are abundant for flat slopes do drift the furthest (see Fig. \ref{fig:results_radii}), but do not lose a lot of their mass starwards of the snowline. The most efficient transport of mass is achieved by meter sized bodies who are most abundant in the -1.83 slope case, where the location at which \SI{50}{\percent} of solids remain in the disk is moved from the snowline to two percent starwards of the snowline.

\subsubsection{Disk influence}
\label{ssec:disk_influence}
\begin{figure}[ht]
\center
\includegraphics[width=\linewidth]{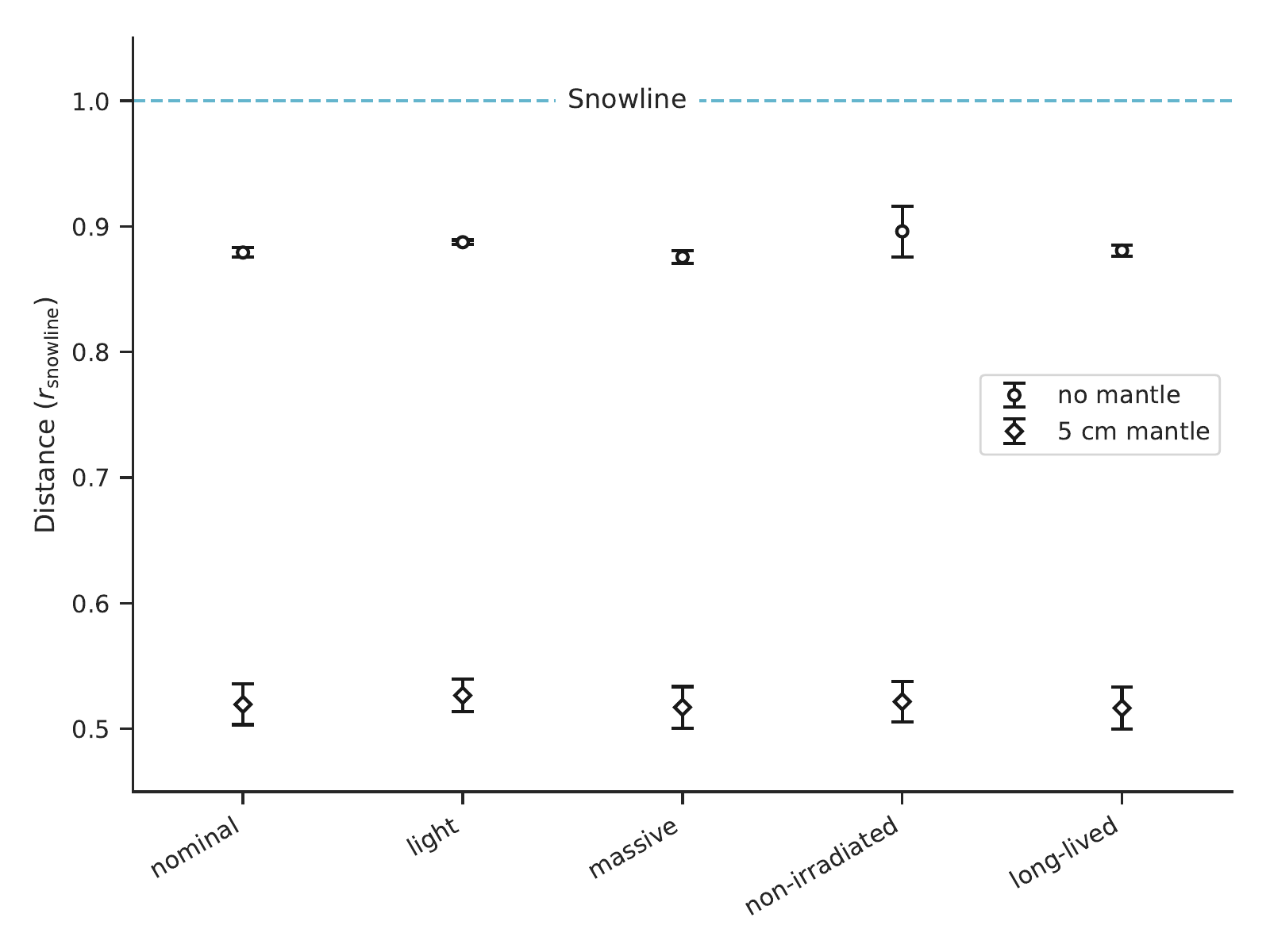}
\caption{Mean relative locations of complete disintegration in different disks for initially \SI{10}{\meter} sized bodies with and without constant dust mantles. As in Fig. \ref{fig:results_radii_rel}, the distance from the central star is measured in units of snowline positions (snowline location at $1$, star at $0$). The mean over the calculations at different times is taken and the $1\sigma$ error is indicated. Refer to the text for the different disk properties.}
\label{fig:results_disks_rel}
\end{figure}

In addition to the nominal disk with values given in table \ref{tab:disk_initial}, we repeated the calculations for bodies with a radius of \SI{10}{\meter} embedded in disks for which we modified one parameter compared to the nominal case: a light disk ($M_\text{disk} = \SI{0.01}{\umass}$, i.e. $\Sigma_0=\SI{53.704}{\gram\per\centi\meter\squared}$), a massive disk ($M_\text{disk}=\SI{0.1}{\umass}$, i.e. $\Sigma_0=\SI{537.046}{\gram\per\centi\meter\squared}$), a long-lived disk ($\dot{M}_\text{wind}=\SI{3e-9}{\umass\per\yr}$), and a disk without heating by irradiation of the central star ($T_l = 0$ in equation \ref{eqn:midplane_temp}). As before, the cometary nucleus model is started multiple times in all the different disk evolution calculations. Initially, the body is separated from the star by a distance \SI{10}{\percent} larger than the classical snowline distance. To cover the full evolution of the disk, the starting times of the individual calculations are scaled with the lifetimes of the different disks. The markers labelled "no mantle" in Fig. \ref{fig:results_disks_rel} show the temporal mean of all these calculations for the different disk cases with indicated standard deviations ($1\sigma$ error-bars). As in Fig. \ref{fig:results_radii_rel}, the distance is measured in units of the classical snowline.

We find, that most of the different tested disks have influences on the locations of complete disintegration in units of classical snowline distances on the percent level only. In Fig. \ref{fig:results_disks_rel} it is shown that different disk masses and lifetimes (controlled by photo-evaporation) do not change the result significantly.

However, the non-irradiated disk has a very different temperature profile once the viscous heating is no longer dominating. Thus, the snowline location is altered to a large extent moving, at late times, very close to the star (i.e. to \SI{0.14}{\au} during the calculation of the latest datapoint). Due to the proximity of the snowline to the star, the slope of the surface density and thus the pressure gradient starts to decrease, hence reduces the drift speed. Therefore, the location of complete disintegration is located closer to the snowline, i.e. only \SI{3}{\percent} below it, which is significantly different from the other cases.

Overall the resulting disintegration locations are robust for the contrasting tested disk cases at many different times. However, the local pressure gradient has a strong influence.

\subsubsection{Dust mantle influence}
\label{ssec:dust_mantle_influence}
As discussed in Sect. \ref{ssec:dust_mantle_formation} the formation of a dust mantle on a cometary nucleus is likely. We assume here the same for a disk-embedded body and quantify its potential influence. An important factor is the size of the body, since the process of dust mantle formation depends on the gravitational force. In general, bigger grains are more easily ejected from small nuclei. We consider here thin dust mantles to initially exist on bodies in the gas disk with radii of \SI{10}{\meter} and evaluate the dust mantle evolution and the influence on the location of complete disintegration. Our model includes dust formation and removal (described in Sect. \ref{ssec:dust_mantle_formation} and \citet{Marboeuf2012}) without cohesive strength (in the following called the "unstable" model).

To estimate the extreme case, where the dust mantle cannot be removed, simulating an infinitely large cohesive strength, we artificially set the dust mantle to a constant thickness (called the "constant" model).

\begin{figure}[ht]
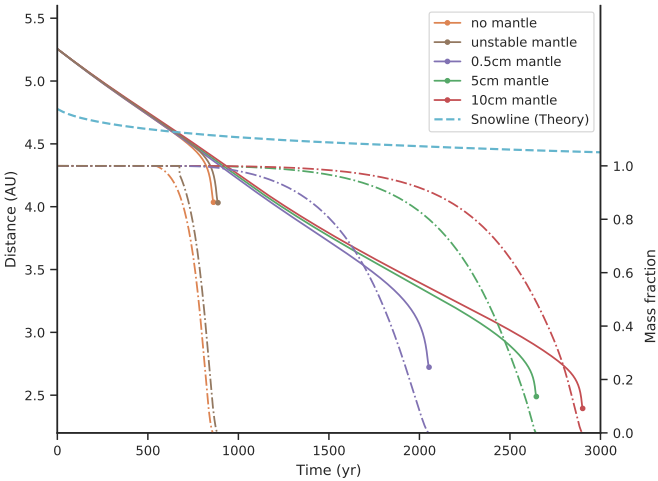

\center
\includegraphics[width=\linewidth]{{{fig_10_results_dust_compare_constant_5_10_05_unstable_5_nominal_disk_mass_initial_mass}.pdf}}
\caption{Sublimation comparison of a \SI{10}{\meter} sized bodies with different dust mantle thicknesses and removal processes. The legend is ordered in increasing sublimation time. The smooth line marks the location of the body in time (left axis), while the dots at the end of the line indicate shrinking to a radius of \SI{10}{\centi\meter} as in Fig. \ref{fig:results_radii} and \ref{fig:results_dust_no_dust}. The dash-dotted lines indicate the mass fraction compared to the initial mass of the same colored case (right axis). The kink that is visible in the mass fraction of the unstable (initially \SI{5}{\centi\meter} thick) mantle stems from the mantle breaking up at that point in time.}
\label{fig:results_dust}
\end{figure}

\begin{figure}[ht]
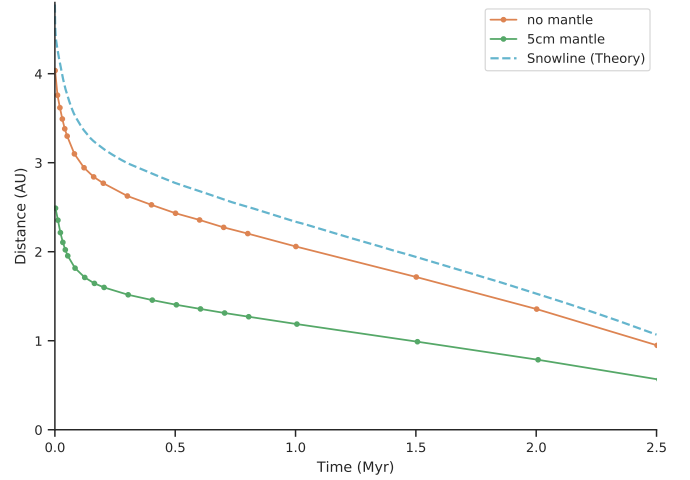

\includegraphics[width=\linewidth]{{{fig_11_results_R10_Sigmanominal_disk}.pdf}}
\caption{Locations of complete disintegration in the nominal disk of initially \SI{10}{\meter} sized bodies with and without a dust mantle. The classical water ice line (snowline) in the disk is indicated for reference. }
\label{fig:results_dust_no_dust}
\end{figure}

In Fig. \ref{fig:results_dust}, it can be seen that using the unstable model the mantle is removed very quickly and the position of complete disintegration differs only slightly from the one without mantle. However, if the dust mantle cannot be removed due to strong cohesive strength and surface sublimation is thus always suppressed, the disintegration location is up to \SI{1}{\au} closer to the star for a \SI{10}{\cm} thick mantle. In Figs. \ref{fig:results_disks_rel} and \ref{fig:results_dust_no_dust}, the less extreme case of a \SI{5}{\cm} thick constant dust mantle is shown.
In the former figure, the temporal mean of the location of complete disintegration for different disks is depicted and in the latter its temporal evolution for the nominal disk is shown.

These results demonstrate the importance of the cohesive strength and thickness of the mantle in determining the thermal evolution of the body. The thickness of the dust mantle is not well constrained by observations, since data is very sparse. The permittivity probe SESAME-PP of the Rosetta mission showed that the first meter is more compact than the rest of the comet 67P \citep{Lethuillier2016}. However, no estimate on the total thickness can be made from this single data point and it is not clear what the composition (possible volatile content) and porosity of this compact layer is. Furthermore, it is not clear how to scale mantle properties from an object with dimensions on the order of kilometer to one with a radius of ten meter.

The large influence of the dust mantle is caused by the change of the sublimation process because free sublimation at the surface is no longer possible if the object is covered by a mantle. Sublimation in the interior still happens, it is however suppressed by the relatively slow diffusion of the released water vapor through the dust mantle, since the small pore radius of the dust mantle limits diffusion.

By analyzing the interior structure of the numerically modeled body, we found that the low thermal conductivity of the dust mantle and porous matrix does not play a dominant role. The body's interior is heated on short timescales on the order of years for the size range we are interested. This can be seen in Fig. \ref{fig:internal_evo_10cm_mantle}, where almost no radial gradient in terms of temperature is visible. This behavior is found for all small body cases with radii of up to \SI{100}{\meter}. For larger bodies or much thicker dust mantles, the picture can change.

The fact that we do not remove the dust mantle by some process is representative of infinite cohesive strength. Thus, the results for a body without dust mantle and the one with constant mantle should be interpreted as lower and upper boundaries for a realistic physical result and the results in Figs. \ref{fig:results_disks_rel} and \ref{fig:results_dust_no_dust} should be interpreted as such. Measuring the distance from the body to the central star in units of snowline distances again, the location of complete disintegration without dust mantle is at $\sim \SI{0.9}{}$, whereas the one with a dust mantle goes down to \SI{0.5} of the snowline distance to the star. However, for a more realistic result, a dust mantle formation and removal model that takes the cohesive strength of the material into account would be needed.

\subsection{Internal thermal evolution}
\label{ssec:internal_evolution}

To analyze the importance of the internal thermal evolution of the body, we first take a look at the results of the comparision of the analytical surface ablation model and the cometary nucleus model (Sect. \ref{ssec:model_comparision}) without a dust mantle.

The underlying assumption of the analytical ablation model is an already equilibrated temperature throughout the body's full structure and the gas. Hence, as expected, the pre-heated cometary nucleus model results are closer to the analytical model. This good agreement between the two models shows that a numerical treatment of the internal evolution is not necessary for bodies composed mainly of dust and water ice with sizes smaller than \SI{100}{\meter} and with initially equilibrated temperatures. For many applications of pure water sublimation it is not necessary to invoke a full model keeping track of the internal structure and temperature because heat conduction - and thus temperature equilibration throughout the body - happens at timescales of years. E.g. for a ten meter sized body, the thermal timescale $\tau_T \simeq R^2  \rho c/K$, where $\rho c$ is the density times the heat capacity and $K$ is the heat conductivity (see table \ref{tab:nucleus_initial}), is approximately \SI{0.3}{\yr}. Thus, the internal temperature of a meter sized body spiraling towards the star on timescales of thousands of years is expected to be in thermal equilibrium with the disk.

\begin{figure}[ht]
	\center
	\includegraphics[width=\linewidth]{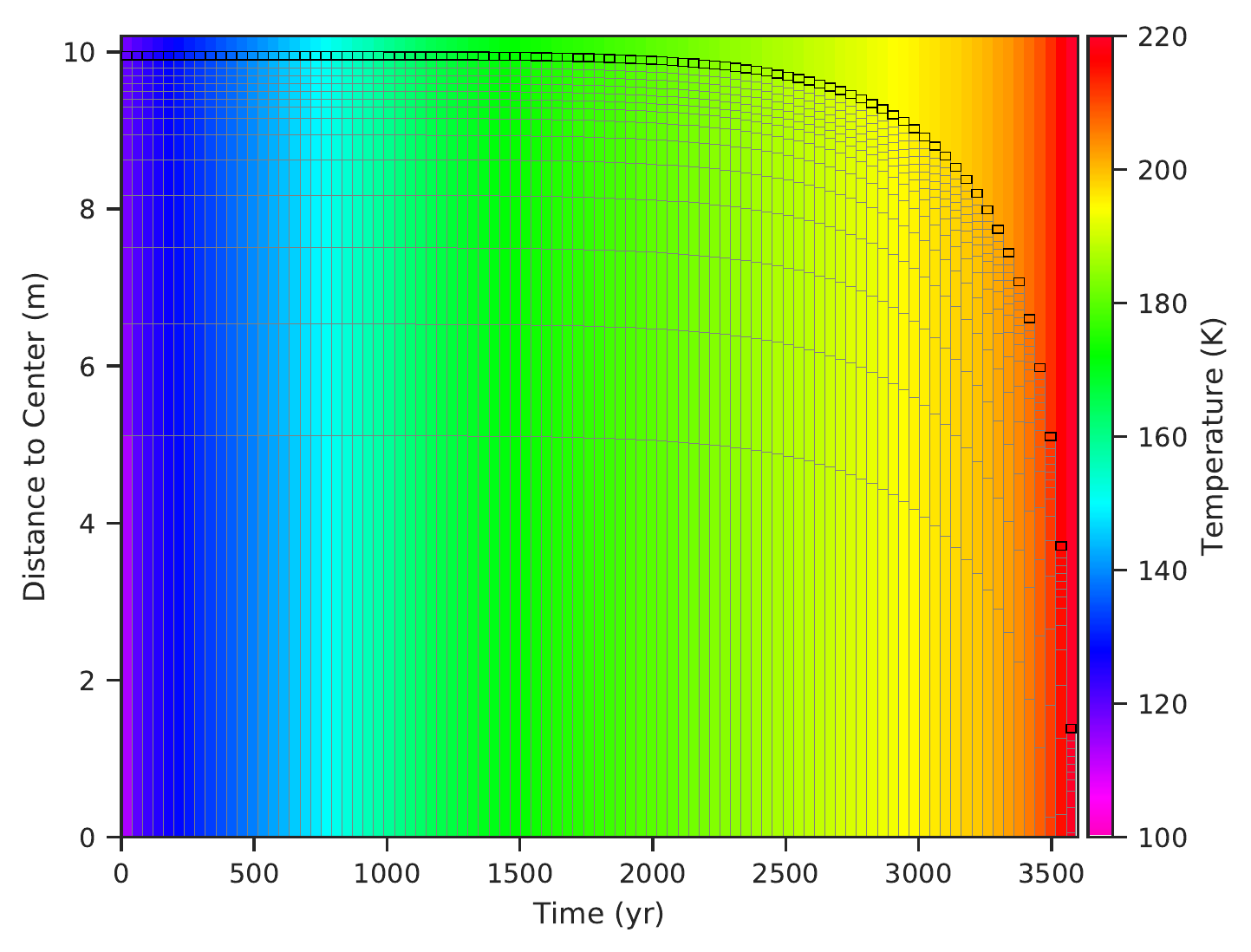}
	\caption{Interior Temperature of an initially \SI{10}{\meter} sized body covered by a \SI{10}{\centi\meter} thick dust mantle. The number of layers is reduced to 15 compared to nominal runs for better visibility and 60 timesteps are merged into one block. The uppermost, dark framed layer shows the dust mantle. Outside the body, the local disk temperature is plotted. A radial temperature gradient is only barely visible.}
	\label{fig:internal_evo_10cm_mantle}
\end{figure}

In the case of a \SI{100}{\meter} sized body within the snowline but without pre-heating, the body shrinks faster than the heat is transported to the interior. However, the cold interior acts as a heat sink. Thus, heat is conducted to the interior which leads to cooler surface temperatures and slower sublimation (see Fig. \ref{fig:comparision_analytical_100m}). This behavior is only reproduced at relatively high temperature regions closer to the central star than the snowline where sublimation is more efficient than heat conduction.

For bodies covered with a dust mantle, the internal temperatures that are reached are significantly higher than for bodies without a dust mantle. Similar is the observed fast heat conduction: for a \SI{10}{\meter} sized body very little variability in the radial direction is visible (Fig. \ref{fig:internal_evo_10cm_mantle}), indicating that sublimation does not lead to a faster shrinking than heat can be conducted to the interior. Thus, the body first becomes isothermal before it disintegrates. No significant thermal insulation increase by the mantle is found: as in the case without a dust mantle, heat conduction acts on timescales of years.

We remark that a numerical treatment of thermal conduction is required to track changes on timescales of years, i.e. on timescales on the order of the orbital period. Hence, assuming an isothermal interior is only valid for objects on almost circular orbits and should not be applied to bodies on eccentric orbits (such as comets).

We conclude that the interior of drifting small bodies ($R<\SI{100}{\meter}$), composed of water ice and dust grains, with zero eccentricity and inclination can be assumed to be isothermal. With that, our analytical model reproduces well the results of the cometary nucleus model. We note that for planetesimals with radii $>\SI{10}{\kilo\meter}$, differentiation due to heating by $^{26}$Al (Sect. \ref{ssec:energy_sources}) can occur \citep{Lichtenberg2016}. For a differentiated body with an ice layer on the surface, sublimation is not hindered by a dust mantle and the analytical sublimation formula becomes appropriate again if no heat is lost to the interior, e.g. if the body is in thermal equilibrium.

\section{Discussion}
\label{sec:discussion}
A number of simplifications and assumptions were made to obtain the presented results. These require discussion and some additional calculations that we describe in this section. In addition to that, a successful test of the radial drift formula is presented in appendix \ref{app:radial_drift_formula_test}.

\subsection{Collisions}
\label{ssec:collisions}
In a protoplanetary disk, collisions are a key evolution factor. In this section, we calculate collision rates between our test body -- called the \textit{target} -- and a population of other bodies present in the disk -- the \textit{impactors} -- and compare them to the timescale of sublimation, which we broadly estimate to be $\sim$\SI{1e3}{\yr}.

Both, the target and the population of impactors undergo radial drift. Even though radial drift timescales can be as short as \SI{1000}{orbital\,periods} \citep{Armitage2019}, they always remain much larger than the orbital period. We calculate collision rates due to two different processes: (a) caused by coupling to the gas (difference in radial and azimuthal velocities) of different sized bodies and (b) due to eccentricity and inclination distributions induced by gravitational stirring and assuming no radial drift. The latter, which we call the {\it orbital collision rate}, is applicable for larger bodies that no longer drift significantly, while the former is applicable for smaller bodies and we call it \textit{Stokes collision rate} to emphasize the coupling to the gas which is quantified by the Stokes number.

In order to compute the collision rates, a statistical approach using a prescribed distribution function of solids is required. For that, the two body (or "particle in a box") approximation \citep{Safronov1969}, i.e. to neglect the influence of the central star, was used to estimate collision rates until \citet{Nakazawa1989a,Nakazawa1989,Ida1989}, and independently \citet{Greenzweig1990,Greenzweig1992} treated collisions using Hills approximation \citep{Hill1878}. The underlying, adopted probability of a test particle hitting a gravitating object (e.g. a planet) during one orbit was derived by \citet{Opik1951}. However, this approach can only be used for the orbital collision rate. For the gas coupled collision rate, we apply the "particle in a box" approach \citep{Safronov1969}. A detailed description of the approaches can be found in appendix \ref{app:collision_rates} and the underlying, assumed mass distribution of bodies as a power law with slope $\alpha$ is described in appendix \ref{app:mass_distribution}.

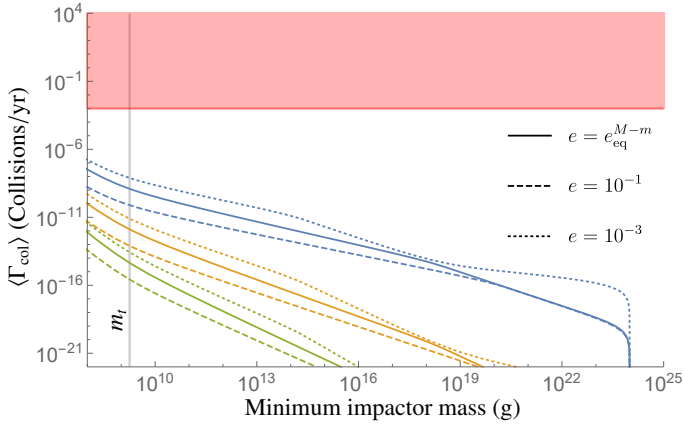
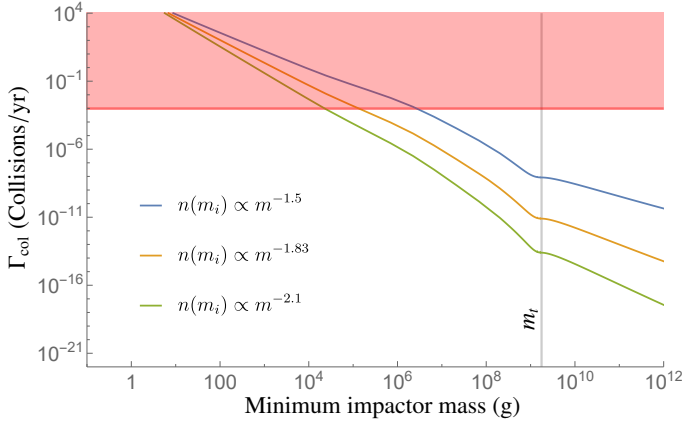
\begin{figure}[ht]
\subfigure[Orbital collision rate\label{fig:Gamma_col_azimuthal}]{\makebox[\linewidth]{
\begin{tikzpicture}
\node (img) {\includegraphics[trim={0 0 0 0},clip,width=0.95\linewidth]{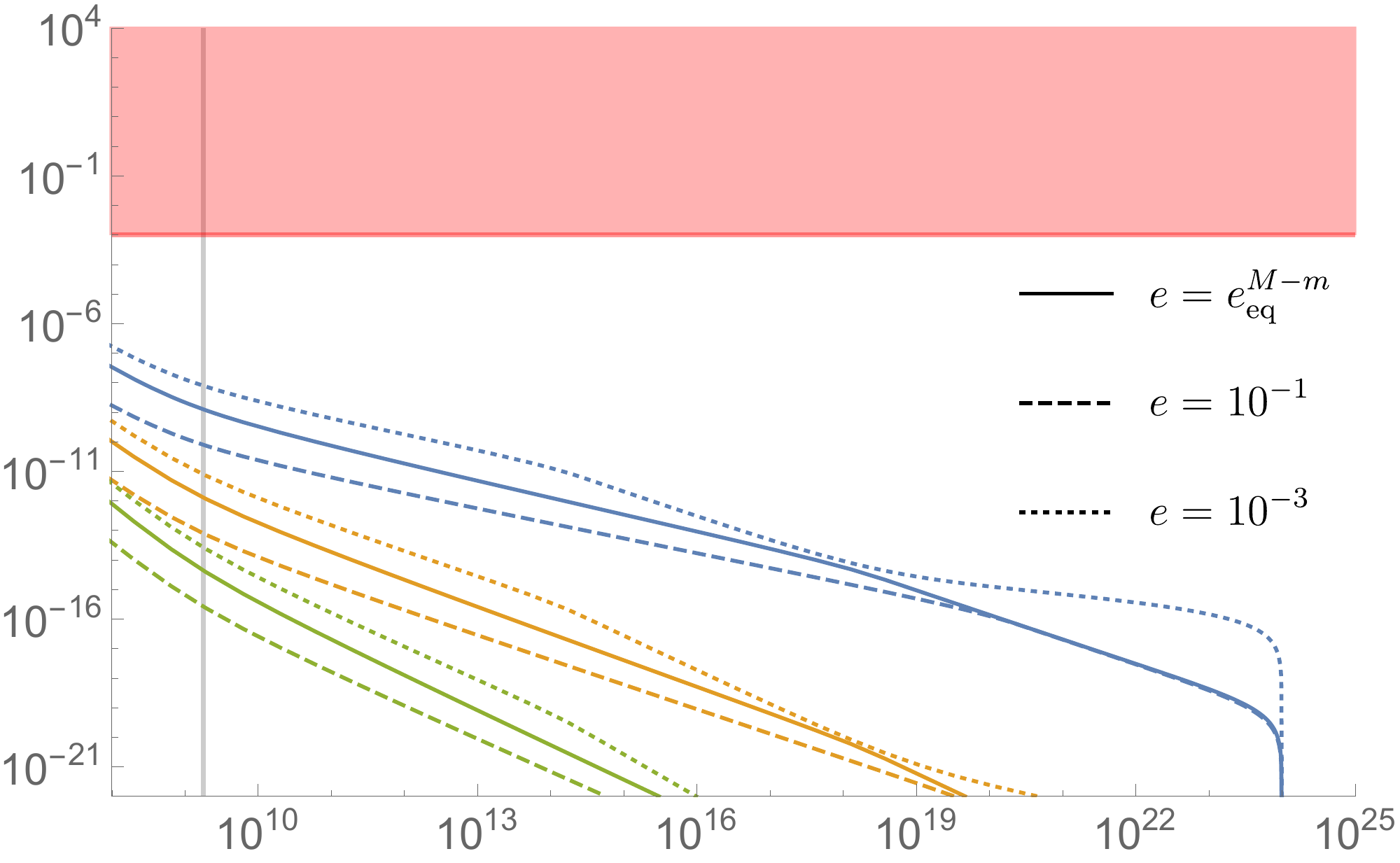}};
\node[below=of img, node distance=0cm, yshift=1.2cm,xshift=+0.3cm]{Minimum impactor mass (\SI{}{\gram})};
\node[xshift=-3.17cm,yshift=-1.6cm,rotate=90]{$m_t$};
\node[left=of img, node distance=0cm, xshift=0.9cm, yshift=1.6cm, rotate=90]{$\braket{\Gamma_\text{col}}$ ($\text{Collisions}/\SI{}{\yr}$)};
\end{tikzpicture}
}}
\subfigure[Stokes collision rate\label{fig:Gamma_col_radial}]{\makebox[\linewidth]{
\begin{tikzpicture}
\node (img) {\includegraphics[trim={0 0 0 0},clip,width=0.95\linewidth]{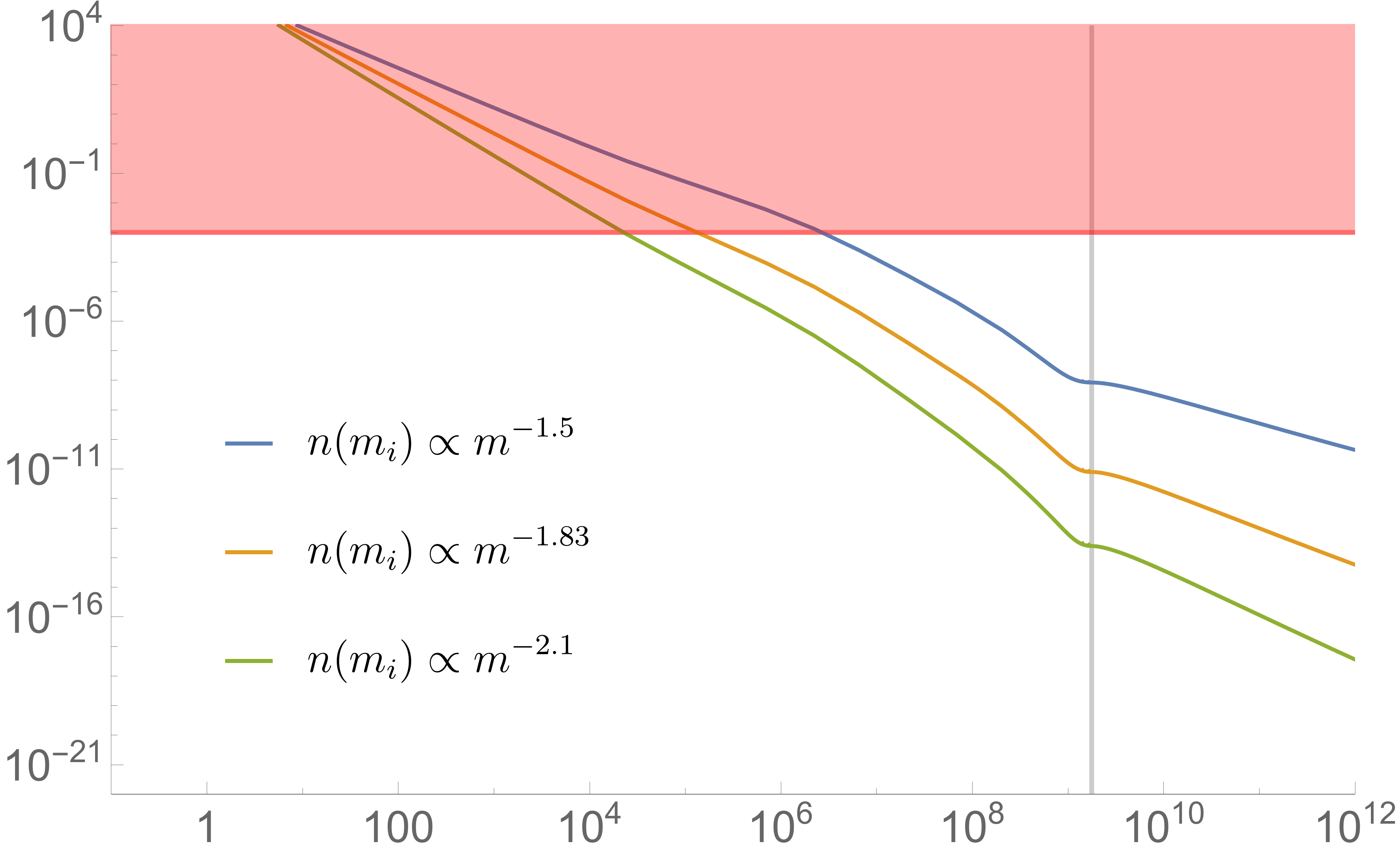}};
\node[below=of img, node distance=0cm, yshift=1.2cm,xshift=+0.3cm]{Minimum impactor mass (\SI{}{\gram})};
\node[xshift=2.25cm,yshift=-1.6cm,rotate=90]{$m_t$};
\node[left=of img, node distance=0cm,xshift=0.9cm,yshift=1.6cm,rotate=90]{$\Gamma_\text{col}$ ($\text{Collisions}/\SI{}{\yr}$)};
\end{tikzpicture}
}}
\caption{Collision rates of the nominal target body (table \ref{tab:nucleus_initial}) with a radius of \SI{10}{\meter} integrated over impactor masses larger than the indicated minimum mass. Results are shown for three different slopes of the impactor mass distribution and in panel (b) additionally for three different eccentricity and inclination values. $e_\text{eq}^{M-m}$ is the equilibrium eccentricity in the vicinity of a mars mass perturber \citep{Ida1990,Thommes2003}. The red shaded region depicts collisions more frequent than once every thousand years and the target mass $m_t$ is indicated.}
\label{fig:Gamma_col}
\end{figure}

The resulting integrated orbital collision rates of our nominal target body with impactors larger than the indicated minimum mass (x-axis) are shown in Fig. \ref{fig:Gamma_col_azimuthal}. Collisions of the target with large impactors $m_i>m_t$ are rare for all considered eccentricity and inclination distributions and are thus negligible. Collisions with smaller bodies have to be treated with the Stokes collision rate prescription (appendix \ref{ssec:radial_coll}) and are shown in Fig. \ref{fig:Gamma_col_radial}. In both cases, the rates were integrated from the minimum impactor mass (on the x-axis) to the maximum mass of \SI{1e24}{\gram}.\footnote{The Stokes collision rates for bodies more massive than the target agree to an order of magnitude precision with the orbital collision rates and do not contribute significantly to the integral.}

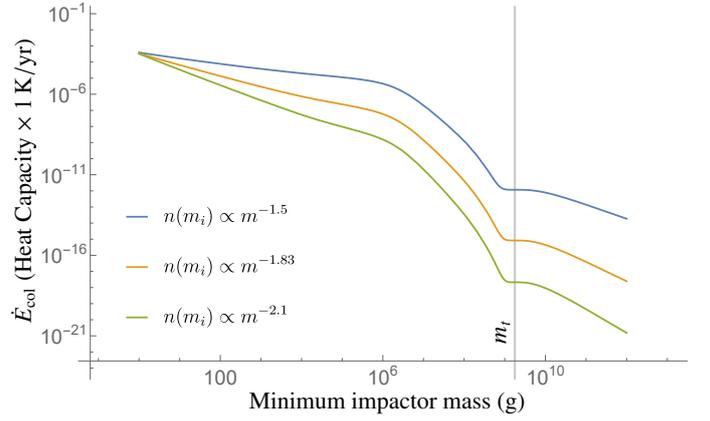
\begin{figure}[ht]
\begin{tikzpicture}
\node (img) {\includegraphics[,clip,width=0.95\linewidth]{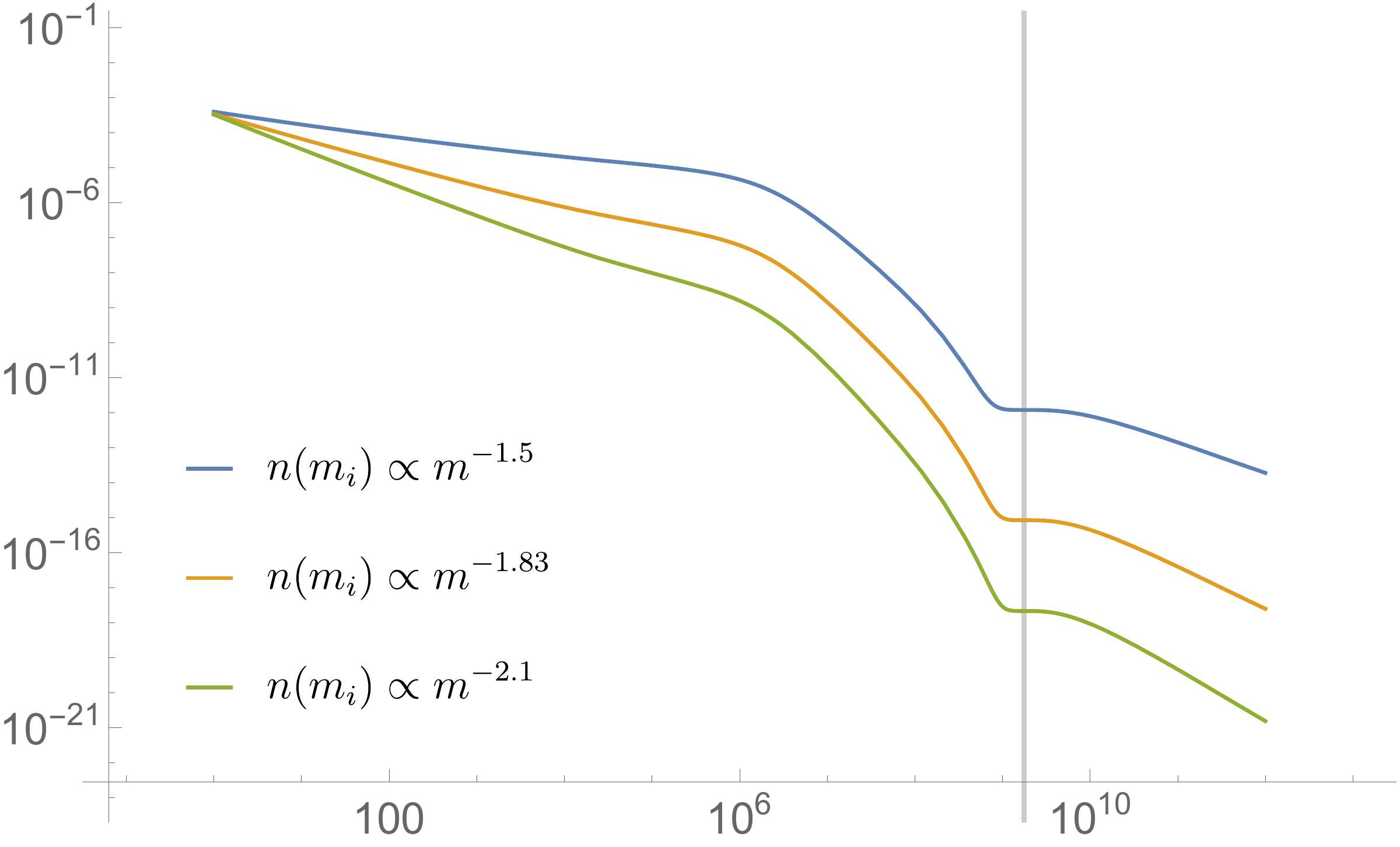}};
\node[below=of img, node distance=0cm,yshift=1.2cm,xshift=0.3cm]{Minimum impactor mass (\SI{}{\gram})};
\node[left=of img,node distance=0cm,xshift=0.9cm,yshift=2.2cm,rotate=90]{$\dot{E}_\text{col}$ (Heat Capacity $\times$ \SI{1}{\kelvin}$/\SI{}{\yr}$)};
\node[xshift=1.82cm,yshift=-1.8cm,rotate=90]{$m_t$};
\end{tikzpicture}
\caption{Collisional energy calculated with the Stokes collision rate, integrated over impactor masses larger than the indicated minimum mass. The energy is measured in units of the energy required to heat the body by one Kelvin. The target properties and impactor mass distributions are the same as in Fig. \ref{fig:Gamma_col} and the target mass $m_t$ is indicated.}
\label{fig:E_col}
\end{figure}

For a very flat mass distribution, relatively high-energy impacts with bodies with diameters larger than \SI{1}{\meter} -- leading to fragmentation \citep{Windmark2012,Blum2018} -- are frequent, i.e. are happening about once per \SI{100}{\yr}, which is comparable to the simulation time of the nominal, drifting body discussed in Sect. \ref{sec:results}. If the mass distribution is steeper, the target is less likely to encounter this kind of collisions.

In terms of energetics, the collisions do not contribute large amounts of energy compared to the thermal energy of the body or the total sublimation energy (see table \ref{tab:nucleus_initial}): Integrating over all sizes, the kinetic energy is $\lesssim \SI{7e6}{\joule\per\yr}$ (using the definition of the "reduced mass kinetic energy" in \cite{Stewart2009}), which is $\sim \SI{4e-4}{\per\yr}$ of the energy required to heat the body by one kelvin and $\sim \SI{3e-7}{\per\yr}$ of the total sublimation energy. The yearly collisional energy deposited on the target by a population of impactors in units of the heat capacity of the target is shown in Fig. \ref{fig:E_col}. For a smaller, \SI{1}{\meter} sized body, the relative numbers increase by almost an order of magnitude. However, the drift and sublimation timescales are also reduced for this smaller body, again resulting in negligible heating by collisions during this stage of the sublimation process.

Most of the energy input results from collisions with bodies with radii smaller than one meter (see Fig. \ref{fig:E_col}). Locally on the targets, the impacts by these smaller bodies are able to erode away target material. This could be the most severe constraint on the applicability of the presented model. The mass encountered per year by the nominal target is $\sim \SI{4e-4}{}$ times its own mass. \cite{Windmark2012} fitted erosion efficiencies based on laboratory experiments for silicate grains. Using their velocity and mass dependent fit \citep[][equation 17]{Windmark2012}, the total eroded mass relative to the target mass is $\sim \SI{8e-2}{\percent\per\yr}$. This would imply that the assumption of a collision free sublimation is only applicable on timescales $\lesssim \SI{10}{\yr}$. Furthermore, for collisions involving impactors with sizes comparable to the target, fragmentation of both objects can happen and only a remnant with mass smaller than the masses of each object remains but Fig. \ref{fig:Gamma_col_radial} shows that this comparable-size case is rare and can be safely ignored. The erosion rates in the regime of collisions with meter-sized bodies is not well studied and applying the fit of \cite{Windmark2012} is therefore an extrapolation with its inherent flaws. Using the lower limit of the erosional prescription for porous icy agglomerates used in \cite{Krijt2015} yields smaller erosion rates $\sim \SI{2e-2}{\percent\per\yr}$ translating to erosion of less than \SI{7.5}{\percent} of the bodies mass over the time where sublimation was active ($T>\SI{150}{\kelvin}$) in the numerical simulations shown in e.g. Fig. \ref{fig:comparision_analytical_10m}.

We conclude, that during the crucial short phase ($\sim \SI{100}{\yr}-\SI{1000}{\yr}$), where sublimation and fast radial drift take place, collisions with small bodies are happening frequently. The results presented in Sect. \ref{sec:results} are only strictly valid if either the surface density of solids is reduced (e.g. by not converting all solids to pebbles, by less efficient settling, or by accumulation of solids in planets), or erosion is less efficient in the relevant mass regime (large uncertainties of extrapolation of laboratory experiments). Otherwise, erosion by collisions could become an additional relevant mass loss mechanism. In terms of thermal energy, collisions do not heat the body, thereby justifying the thermal balance model we presented in Sect. \ref{sec:model}. The fast erosion of meter sized bodies is the main argument against their presence in disks. In this work, however, we postulate their presence, which could be justified by frequent enough fragmentation of larger bodies.

We note, that the retention of a dust mantle is very hard to achieve if collisions are eroding away the uppermost layers of the body. A mantle of centimeter thickness is eroded by collisions with pebbles in less than \SI{10}{\yr}.

\subsection{Gas versus surface temperatures}
\label{ssec:gas_v_surf_t}
\citet{DAngelo2015} showed that for small bodies ($R<\SI{10}{\kilo\meter}$) the bulk temperature of the body is in equilibrium with the gas after less than 500 orbits (see Fig. 20 in \citet{DAngelo2015}). In their work, the entire body was heated and reached equilibrium temperature in this amount of time, whereas in our full model, we only assume equilibration of the temperature in an uppermost, thin layer. Instantaneous heat exchange from the thermal bath, i.e. the disk, to the body is thus well justified. 

\subsection{Frictional heating}
\label{ssec:frictional_heating}
In this work, heating due to interactions with the non-Keplerian gas is not taken into account. \citet{DAngelo2015} calculate the equilibrium value of the surface temperature for their planetesimals to be \citep[][equation 38]{DAngelo2015}
\begin{equation}
(T_s^\text{eq})^4 \approx T_g^4 + \frac{C_D \rho_0}{32 \sigma \varepsilon_s} |v_g - v_\text{K}|^3\,,
\end{equation}
where $C_D$ is the drag coefficient, $\sigma$ is the Stefan-Boltzmann constant and $\varepsilon_s$ is the thermal emissivity (for a black body $\varepsilon_s = 1$). To derive this equilibrium value, a fraction of $C_D/4$ of the total collisional energy is assumed to be transmitted as heat to the body, which corresponds to an upper limit \citep{Podolak1988}.

For a simple estimate, using the typical values $\rho_0=\SI{e-9}{\gram\per\centi\meter\cubed}$, $T_g=\SI{140}{\kelvin}$ and $\eta=\SI{4e-3}{}$, frictional heating yields a negligibly small temperature increase of \SI{8e-4}{\kelvin} for a black body. Only in very dense regions of the disk or potentially in the atmospheres of planets could a significant change occur.

\subsection{Water vapor pressure}
\label{ssec:water_vapor_pressure}
Disk water vapor can change the sublimation rate and even lead to deposition of water onto bodies if present in high enough abundance ($P_\text{vapor}>P^s(T)$). In an ideal case, where all other solid bodies in the disk do not move radially and the disk is not evolving, $P_\text{vapor}<P^s(T)$ everywhere. Hence, no water would condense onto the surface of a body drifting by. However, if fast drifting pebbles are present, the water vapor surface density can be replenished by diffusion of the freshly released vapor starwards of the snowline \citep{Ros2013,Schoonenberg2017,Drazkowska2017}. Another source for out of thermal equilibrium water vapor could potentially be stellar outbursts which episodically heat up the disk \citep{Hartmann1996}. To study constraints for deposition or suppressed sublimation in detail, a model including the evolution of all solids and water vapor in the disk would be needed. If a significant amount of vapor is transported further away from the star than the snowline, it could be deposited onto a drifting body, reducing (for Stokes numbers $s > 1$) the drift speed, which allows for even more deposition, potentially leading to growth to planetesimal size.

A local source for enhanced water vapor pressure could also be the drifting body itself, due to exhibiting a coma-like region with enhanced partial pressure of water, reducing the sublimation rate. The transport of gas or vapor away from the body is not modeled here and would differ from the case of a comet due to the disk gas interacting with the released vapor and the lack of solar wind, which is absorbed in the disk. Our assumption of no increased local partial pressure due to the coma is consistent with a complete erosion of the coma by the disk gas.

\begin{figure}[ht]
\center
\includegraphics[trim={0cm 0.733cm 0cm 0.16cm},clip,width=\linewidth]{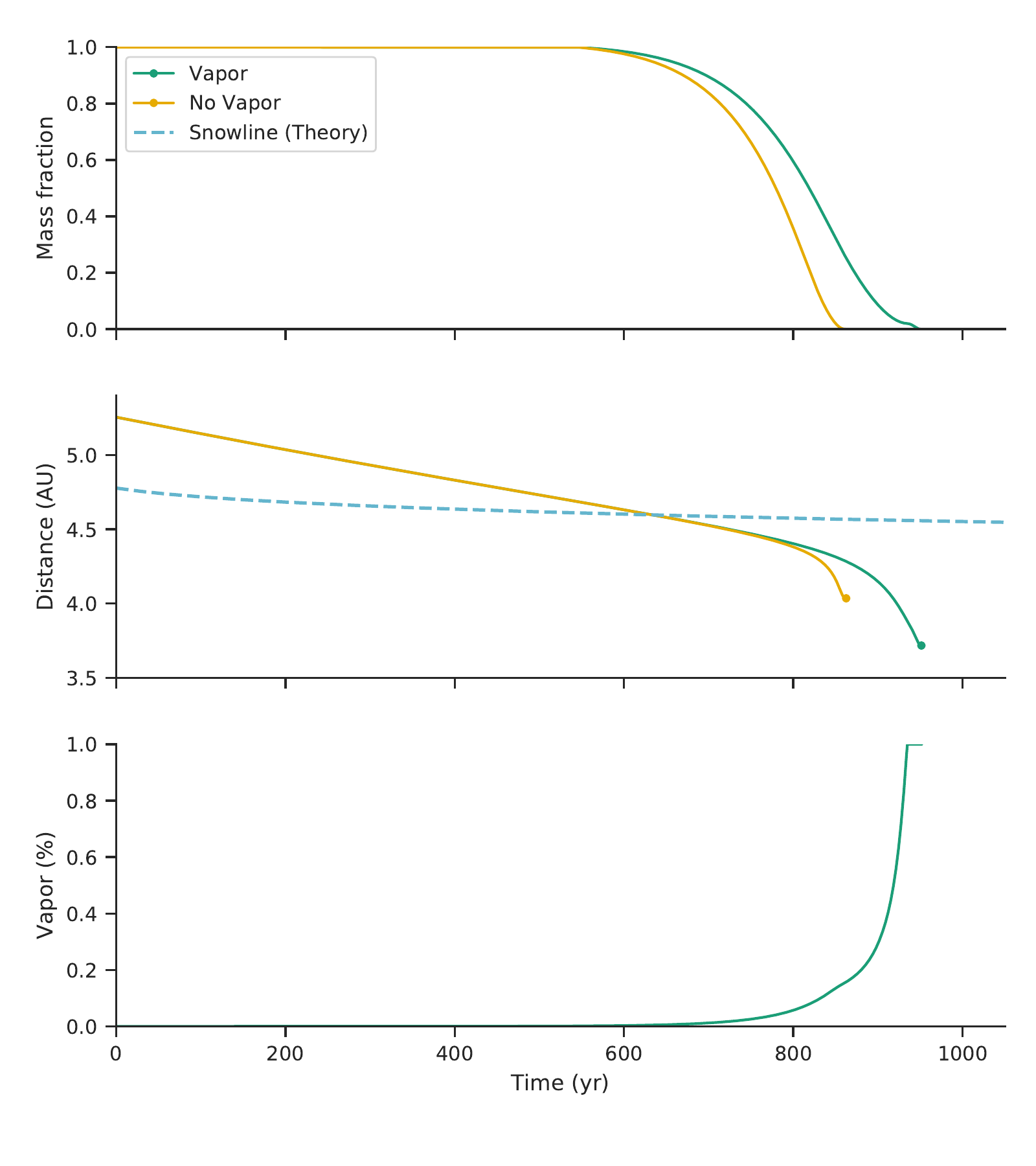}
\caption{Evolution of a \SI{10}{\meter} sized body in the nominal disk calculated with the analytical surface ablation model, with and without water vapor pressure. The water vapor increases exponentially depending on the local disk temperature up to a maximum of one percent of the local pressure.}
\label{fig:vapor_compare}
\end{figure}

If the partial pressure of vapor is smaller than the sublimation pressure $P^s$ but not zero, it reduces the sublimation rate compared to the nominal results without vapor. In Fig. \ref{fig:vapor_compare} we show the influence of an artificially chosen, exponential increase of partial pressure of water vapor (motivated by results of the pebble based evaporation models of \citet{Schoonenberg2017}) up to one percent of the total pressure. The location to reach the percent level is set to where the temperature is \SI{176.6}{\kelvin} to avoid deposition of water.

Under these assumptions, the water vapor moves the location of complete disintegration farther in towards the star. Hence, in the context of presence of water ice in solid bodies (the "dynamical" snowline), the results for the case without a mantle and without water vapor in the disk is an upper boundary for the dynamical snowline location.

\section{Conclusions}
\label{sec:conclusions}
We presented the application of a cometary nucleus model to disk-embedded, radially drifting, spherical bodies, tracking the thermodynamic evolution of the object. The body is assumed to consist of only dust and water ice. Different properties of the disks and the drifting bodies were explored and the time evolution of the disk was taken into account. The main focus of the work was to constrain the regions that can be reached by drifting icy bodies, ultimately determining the zone where some water can be incorporated in solids and thus be accreted by growing terrestrial planets.

Here, we summarize the key findings:  
\begin{enumerate}
\item Almost independently of the properties and temporal evolution of the disk, drifting bodies with radii $\ge \SI{1}{\meter}$ can transport water ice at least ten percent closer to the star than the location of the "classical" snowline before they completely disintegrate.
\item If surface sublimation is not impeded in any way, e.g. by the presence of a dust mantle, it is the dominant process for the evolution of the object and can be modeled in a simple, analytic way with good agreement with the results of a full numerical model.
\item These results are applicable to bodies with radii ranging from meters to \SI{100}{\meter}. Smaller bodies never experience fast radial drift, therefore the effect is suppressed, whereas bodies larger than \SI{100}{\meter} do not drift fast enough to even cross the snowline. In the range from tens to hundreds of meters, the difference in locations of complete disintegration is small. This implies that if bodies in this size range are present, a quantifiable smearing of the water snowline results. In the absence of meter-sized bodies, the snowline is given by the local disk properties only.
\item A dust mantle covering the body suppresses surface sublimation and forces the internally released vapor to diffuse through the mantle. For the extreme case of a non-breakable mantle, this results in icy bodies drifting starwards to about one half of the classical snowline position. However, the presence and formation of a global dust mantle on a body embedded in a protoplanetary disk is hindered by collisions with pebble sized objects, because these collisions occur at relative velocities typically leading to net mass loss, i.e. erosion of the uppermost layers of the body. In particular, for bodies smaller than meter-size a dust mantle is highly unlikely to be kept due to the -- relative to the total mass -- large erosion rates.
\end{enumerate}

Multiple processes were not included and several assumptions were made to obtain the above results. We identified two key processes that could affect our results and which should be addressed in future works:
\begin{itemize}
\item Collisions with different sized bodies, mainly stemming from the difference in radial and azimuthal velocities due to gas drag, are frequent for large solid fractions and would mostly lead to erosion. For a model including multiple bodies of different sizes, tracking the thermodynamic evolution of each is necessary to properly estimate the general outcome. A potential approach to reduce the numerical cost is to use the analytic surface sublimation expression for objects with low thermal variability over the course of an orbit (i.e. low eccentricity and inclination).
\item Specific water vapor pressures influence sublimation rates and thus the results are sensitive to this. To get fully consistent results, it is required to take the water vapor distribution in the disk, including the contribution of the evaporation bodies, into account.
\end{itemize}
The approximations of imposing the gas temperature as surface temperature, neglecting frictional heating, and using a simple formula for the radial drift is found justified for all discussed parameters.

Of particular interest for future works is to test and potentially apply the analytic sublimation formula in complete N-body terrestrial planet formation models \citep[as suggested by][]{Coleman2016}. This would also include larger than \SI{100}{\meter} sized bodies because they could be moved across the snowline by N-body interactions (e.g. scattering or resonant trapping) and bodies on significantly eccentric and inclined orbits for which further research on their thermal evolution is necessary. Furthermore, we did not include different chemical species that could either be present as icy layers on the grains or as clathrates and we leave the treatment of the evolution of bodies at different, potentially observable ice lines to future works. Moreover, the influence of including amorphous water ice and the phase change to crystalline ice along with a model for dust mantle growth including cohesive strength and predicting the properties (pore size, porosity, tortuosity, thickness) of the formed mantle should also be addressed in the future for a complete model.

The presented and proposed steps will help to constrain compositions and available masses for terrestrial planet growth, which will be increasingly required to match the precisions on future observational constraints on planetary compositions.

\acknowledgements
The authors thank the anonymous referee who's comments helped improve the manuscript and Cléa Serpollier for her early investigative work in the subject.
This work has been carried out within the frame of the National Centre for Competence in Research PlanetS funded by the Swiss National Science Foundation (SNSF). The authors acknowledge the financial support from the SNSF under grant 200020\_172746.

\bibliography{library}{}

\appendix
\section{Radial drift formula}
\label{app:radial_drift_formula_test}
Treating the fastest drifting bodies in protoplanetary disks correctly, might require additional changes to the radial drift formula shown in equation \eqref{eq:radial_drift}. We show here the validity of the assumptions made to derive this form of the equation, i.e. assuming orbit averaged drift $\tau_\text{drift} \gg \tau_\text{orb}$, neglecting terms quadratic in $\eta$, assuming no radial acceleration ($dv_{r,s}/dt=0$), and setting the particle's azimuthal speed to Keplerian  ($v_{\theta,s}=v_\text{K}$) in the derivative term (first term in equation  \eqref{eqn:second_part_eom_rad_drift})\footnote{See \cite{Takeuchi2002} for an instructive derivation of the simplified equations}. The equations of motion in the disk plane ($v_z = 0$) are \citep{Takeuchi2002}
\begin{align}
\frac{d v_{r,s}}{dt}=\frac{v_{\theta,s}}{r}-\Omega_\text{K}^2 r - \frac{\Omega_\text{K}}{t_\text{stop}}\left(v_{r,s}-v_{r,g}\right)\,,\\
\frac{d}{dt}\left(rv_{\theta,s}\right)=-\frac{v_\text{K}}{t_\text{stop}}\left(v_{\theta,s}-v_{\theta,g}\right)\,,
\label{eqn:second_part_eom_rad_drift}
\end{align}
where the subscript $s$ and $g$ are for the solid body and gas, respectively, and $t_\text{stop}$ is given by equation \eqref{eq:stopping_time}.
\begin{figure}[ht]
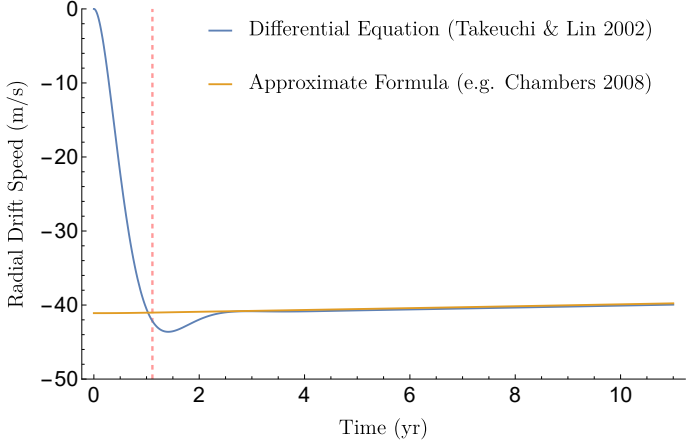

\center
\includegraphics[width=\linewidth]{{{radial_drift_diff_plot_s100_T140K}.pdf}}
\caption{Radial drift speed comparision between the approximate equation \ref{eq:radial_drift} and the numerical solution to the differential equations. The radius is \SI{1}{\meter}, the initial location is at \SI{2}{\au}, temperature and midplane density are constant over the disk and set to \SI{140}{\kelvin} and \SI{2.5e-10}{\gram\per\centi\meter\cubed}. Initially the body moves with Keplerian speed in azimuthal direction and no radial velocity. The dashed, vertical line marks the stopping time.}
\label{fig:radial_drift_test}
\end{figure}
For a test, we assumed $v_{r,g}=0$ and solved the equations of motion numerically. The results can be seen in Fig. \ref{fig:radial_drift_test} and are compared to the results of the analytical equation \eqref{eq:radial_drift} with the same initial conditions. After one stopping time has passed (dashed vertical line), the initially Keplerian azimuthal ($v_{\theta,s}(t=0)=v_\text{K}$) speed slowed down to an equilibrium value and the analytical expression \eqref{eq:radial_drift} reproduces the differential equation results well. The radial drift speed is slowing down because the body moves towards the star ($dr/dt \propto r \eta(r)$). The order of percent difference after equilibration of the azimuthal speed can thus be explained by this non-zero $d v_{r,s}/dt$, which is assumed to be zero to derive equation \eqref{eq:radial_drift}.This difference is small compared to the uncertainties of the other processes treated in this work.

\section{Mass distribution}
\label{app:mass_distribution}
Before assessing the collision rates, we discuss here briefly the required mass or size distributions of the bodies in the disk. The differential mass distribution $n(m_i)$ is defined such that $n(m_i)dm_i$ is the number of bodies with masses in the interval [$m_i$,$m_i+dm_i$]. We describe $n(m_i)$ as a power-law with exponent $\alpha$. Data constraining the mass distribution is mainly available from solar system observations or from theoretical works treating collisional cascades or related effects \citep[e.g.][]{Dohnanyi1969,Tanaka1996,Makino1998,Benz1999,Jutzi2010,Pan2012,Belton2015}. The observational data is either gathered by direct measurements of Jupiter family comets \citep[e.g][]{Fernandez1999,Tancredi2006,Fernandez2013}, trans-Neptunian objects \citep[e.g.][]{Bernstein2004} or asteroids \citep[e.g.][]{Gladman2009} or inferred from distributions of craters on planets, satellites or other minor planets \citep{Zahnle2003,Singer2019}. The measured and predicted values of the slope $\alpha$ of the differential mass distribution -- assuming a fixed density for converting size distributions -- lie in the interval $[-1.5,-2.1]$. We note that multiple studies found different slopes for bodies with radii smaller than \SI{}{\kilo\meter} \citep{Zahnle2003,Fernandez2006,Fernandez2013,Singer2019}. Nevertheless, we adopt for our order of magnitude estimates simple unbroken power laws with three fixed values for $\alpha$: the upper \citep[-1.5 as][]{Morbidelli2015} and lower \citep[-2.1 slightly lower than the -2.05 found by][]{Belton2015} limits and the $\alpha$ resulting from the self-similar solution to the collisional cascade \citep[-1.83,][]{Dohnanyi1969}.

To avoid divergence, the distribution needs to be cut at a lower and an upper boundary. We choose an upper limit to the mass of \SI{1e24}{\gram}, corresponding to a radius of \SI{827}{\kilo \meter}. The lower cut is of particular importance for the resulting collisions rates. We choose the typical pebble that can form by coagulation for the lower limit: according to laboratory experiment it has a size of $\sim$\SI{1}{\centi\meter} and a corresponding mass of $\sim$\SI{1}{\gram} \citep{Blum2018}. To not underestimate the amount of solids, we assume a \SI{100}{\percent} conversion of dust to pebbles and larger bodies.

\section{Collisions}
\label{app:collision_rates}

\subsection{Orbital collision rate}
\label{ssec:azimuthal_coll}
The averaged number of collisions between a target with mass $m_t$ and a population of bodies with mass $m_i$ per unit time is written as \citep{Nakazawa1989a,Ohtsuki1999,Inaba2001}
\begin{equation}
\braket{\Gamma_\text{col}}_{ti} = h^2_{ti}a^2\Omega_K n_s(m_i)dm_i \braket{P_\text{col}}_{ti}\,,
\label{eq:gamma_col}
\end{equation}
where $\braket{P_\text{col}}_{ti}$ is a non-dimensional mean collision rate between bodies with masses $m_t$ and $m_i$ that is independent of the total number of bodies, but depends on the common semi-major axis, the radii and masses of the two bodies, and the mass of the central star. The brackets indicate, that the mean collision rate is an average over all eccentricities and inclinations given by a Reyleigh-type distribution function with eccentricity (inclination) dispersions $e^*$ ($i^*$), which also influence the mean collision rate \citep{Inaba2001}. $n_s(m_i)dm_i = \Sigma_s/m_i\, n(m_i) dm_i$ is the surface number density of bodies with masses between $m_i$ and $m_i+dm_i$ with $\Sigma_s$ the surface density of solids, whereas $h_{ti}$ is the reduced Hill radius of two bodies with masses $m_t$ and $m_i$ given by
\begin{equation}
h_{ti}=\left(\frac{m_t+m_i}{3M_*}\right)^{1/3}\,.
\end{equation}

For the entire range of realistic eccentricity and inclination distributions, \citet{Inaba2001} found that numerical results are well reproduced if the non-dimensional mean collision rate is set to 
\begin{equation}
\label{eq:P_col}
\braket{P_\text{col}}=\min\left( \braket{P_\text{col}}_\text{med},\left(\braket{P_\text{col}}^{-2}_\text{high}+\braket{P_\text{col}}^{-2}_\text{low}\right)^{-1/2}\right)\,,
\end{equation}
where the individual parts are
\begin{itemize}
\item \begin{equation}
\label{eq:P_col_high}
\braket{P_\text{col}}_\text{high} = \frac{\tilde{r}_p^2}{2\pi} \left(\mathcal{F}(I^*)+\frac{6}{\tilde{r}_p}\frac{\mathcal{G}(I^*)}{({{\tilde{e}}^*})^2}\right)\,,
\end{equation}
where $I^* \equiv \tilde{i}^*/\tilde{e}^*$,
\begin{align}
\mathcal{F}(I^*) \equiv  8 \int_0^1 \frac{{I^*}^2 E[\sqrt{3(1-\lambda^2)}/2]}{[{I^*}^2+(1-{I^*}^2)\lambda^2]^2} d\lambda\\ \intertext{and}
\mathcal{G}(I^*) \equiv  8 \int_0^1 \frac{K[\sqrt{3(1-\lambda^2)}/2]}{[{I^*}^2+(1-{I^*}^2)\lambda^2]}d\lambda\,,
\end{align}
where $K$ and $E$ are the complete elliptic integrals of the first and second kinds,
\item \begin{equation}
\braket{P_\text{col}}_\text{med} = \frac{\tilde{r}_p^2}{4\pi \tilde{i}^*} \left(17.3+\frac{232}{\tilde{r}_p}\right)
\end{equation}
\item \begin{equation}
\braket{P_\text{col}}_\text{low} = 11.3\sqrt{\tilde{r}_p}\,,
\end{equation}
\end{itemize} 
with the reduced eccentricity and inclination dispersions
\begin{equation}
\tilde{e}^* \equiv e^*/h_{ti}\,,\quad \tilde{i}^*\equiv i^*/h_{ti}
\end{equation}
and
\begin{equation}
\tilde{r}_p \equiv \frac{R_t+R_i}{h_{ti}a}\,.
\end{equation}

\subsection{Stokes collision rate}
\label{ssec:radial_coll}
For drifting, small particles we use the "particle in a box" approximation, where the collision rate $\Gamma_\text{col,ti}$ of a gravitating target with radius $R_t$ and with a number of impactors with radius $R_i$ is \citep{Safronov1969}
\begin{equation}
\Gamma_\text{col,ti} = n_V(m_i) \pi (R_t+R_i)^2 \Delta v \left(1+\frac{v_\text{esc}^2}{\Delta v^2}\right)\,,
\end{equation}
with the volume number density of impactors $n_V(m_i)$, the relative velocity $\Delta v$ of the impactors with respect to the target and the mutual escape speed
\begin{equation}
v_\text{esc}^2= 2G\frac{m_t+m_i}{R_t+R_i}\,.
\end{equation}
To estimate the collision rate, we use the squared sum of the difference in radial (equation \eqref{eq:radial_drift}) and azimuthal drift velocities of the target and impactors according to their size. The azimuthal difference in velocity is given by $\eta v_k | \frac{1}{1+s_i^2}-\frac{1}{1+s_t^2}|$, where $s_i$ and $s_t$ are the Stokes numbers of the impactor and the target \citep{Birnstiel2016}. This excludes the additional velocity components due to Brownian motion and turbulence. For a more complete discussion of relative velocities we refer to \cite{Ormel2007}.

The number density of a given size of impactors can be estimated given three ingredients: their mass (or size) distribution discussed above, the density of the gas and the local dust to gas ratio $f_\text{solid}$, which is locally enhanced due to dust settling and can for low masses be described by \citep{Youdin2007,Birnstiel2016}
\begin{equation}
f_\text{solid}=\frac{0.01}{\sqrt{\frac{\alpha_Z}{s+\alpha_Z}}}\,,
\end{equation}
where $s$ is the Stokes number and $\alpha_Z$ is a dimensionless parameter for turbulent diffusion in the vertical direction. Here, we assumed a global dust to gas fraction of \SI{0.01}{} and we assume $\alpha_Z \simeq \alpha$, which is true on orders of magnitude level \citep{Youdin2007}. For larger than meter-sized bodies the settling is no longer well described by the processes considered in \cite{Youdin2007}, since gravitational interactions between the particles become important. Therefore $f_\text{solid}$ is restricted to be $f_\text{solid}\leq1$ and this maximum is reached at $R_2\approx\SI{5}{\meter}$. At meter size, the inclination caused by the viscous stirring of a larger planetesimal in the vicinity of our target leads to approximately the same elevation above the midplane as the reduced scale height \citep{Ida1990,Thommes2003,Fortier2013}.

\end{document}